\documentclass[aps,prd,10pt,twocolumn,superscriptaddress,floatfix,nofootinbib,amsmath,amssymb,altaffilletter,preprintnumbers]{revtex4-1}

\usepackage{graphicx}
\usepackage{dcolumn}
\usepackage{bm}
\usepackage{tensor}
\usepackage{slashed}
\usepackage{bm}
\usepackage{multirow}
\usepackage{soul}
\usepackage{amsmath}
\usepackage{mathrsfs}
\usepackage{amssymb}
\usepackage{yfonts}
\usepackage{color}
\usepackage{xspace}
\usepackage{url}
\usepackage{verbatim}
\usepackage{mathtools}
\usepackage{upgreek}
\usepackage{amstext}
\usepackage{booktabs}
\usepackage{tabulary}
\usepackage{tabularx}
\usepackage{etoolbox}
\usepackage[utf8]{inputenc}
\usepackage[dvipsnames,table]{xcolor}
\newcolumntype{Y}{>{\centering\arraybackslash}X}
\usepackage[colorlinks=true,pagebackref=true,pdfstartview=FitV,breaklinks=true]{hyperref}
\hypersetup{urlcolor=BlueViolet,
	    citecolor=Plum,
	    linkcolor=PineGreen}
\usepackage[official]{eurosym}
\definecolor{lightgray}{rgb}{0.9,0.9,0.9}	    
\definecolor{green}{rgb}{0,0.5,0}
\definecolor{red}{rgb}{1,0,0}
\definecolor{blue}{rgb}{0,0,0.5}

\newcommand{\dbd}[2]{\ifmmode \frac{\textrm{d}#1}{\textrm{d}#2}\else $\textrm{d}#1/\textrm{d}#2$\fi}
\newcommand{\pbp}[2]{\ifmmode \frac{\partial#1}{\partial#2}\else $\partial#1/\partial#2$\fi}

\newcommand{\drm}{\mathrm{d}}

\DeclareMathAlphabet{\mathpzc}{OT1}{pzc}{m}{it}
\newcommand{\eV}{\text{e\kern-0.15ex V}\xspace}

\newcommand{\TeV}{\text{T\kern-0.1ex \eV}\xspace}

\newcommand{\cevns}{CE$\nu$NS\xspace}
\newcommand{\hecf}{He:CF$_4$\xspace}
\newcommand{\cffour}{CF$_4$\xspace}

\newcommand{\keVr}{\text{k\text{e\kern-0.15ex V}$_\mathrm{r}$}\xspace}

\newcommand{\Cygnus}{\textsc{Cygnus}\xspace}

\newcommand{\be}{\begin{equation}}
\newcommand{\ee}{\end{equation}}
\newcommand{\bea}{\begin{eqnarray}}
\newcommand{\eea}{\end{eqnarray}}

\begin{document}

\title{Directional recoil detection for CEvNS measurements with light nuclei at the Spallation Neutron Source}

\author{Ciaran A.~J.~O'Hare}
\email{ciaran.ohare@sydney.edu.au}
\affiliation{ARC Centre of Excellence for Dark Matter Particle Physics, The University of Sydney, School of Physics, NSW 2006, Australia}

\author{Anirudh Chandra Shekar}
\affiliation{Department of Physics and Astronomy, Mitchell Institute for Fundamental Physics and Astronomy,
Texas A$\&$M  University, College Station, Texas 77843, USA}

\author{Chiara Lisotti}
\affiliation{ARC Centre of Excellence for Dark Matter Particle Physics, The University of Sydney, School of Physics, NSW 2006, Australia}

\author{Michael Litke}
\affiliation{Department of Physics and Astronomy, University of Hawaii, 2505 Correa Road, Honolulu, HI, 96822, USA}

\author{Nityasa Mishra}
\affiliation{Department of Physics and Astronomy, Mitchell Institute for Fundamental Physics and Astronomy,
Texas A$\&$M  University, College Station, Texas 77843, USA}

\author{Jayden~L.~Newstead}
\affiliation{ARC Centre of Excellence for Dark Matter Particle Physics, School of Physics, The University of Melbourne, Victoria 3010, Australia}

\author{Louis E. Strigari}
\affiliation{Department of Physics and Astronomy, Mitchell Institute for Fundamental Physics and Astronomy,
Texas A$\&$M University, College Station, Texas 77843, USA}

\author{Sven E. Vahsen}
\affiliation{Department of Physics and Astronomy, University of Hawaii, 2505 Correa Road, Honolulu, HI, 96822, USA}

\smallskip
\begin{abstract}
The coherent elastic scattering of neutrinos on nuclei, also known as \cevns, has been studied for several years by the COHERENT program of experiments using neutrinos from stopped-pion decays produced at the Spallation Neutron Source (SNS). We propose a new approach for \cevns measurements at the SNS that aims to complement the COHERENT experiments in two main ways: by reconstructing the \textit{angular} distribution of \cevns-induced recoils, and by measuring \cevns on much lighter target nuclei such as helium, carbon, and fluorine. The proposed detector would employ a gaseous time-projection chamber with a highly segmented charge readout to enable the spatial reconstruction of $\sim$10--500 keV ionisation tracks created by \cevns-induced recoils. This would enable the simultaneous measurement of the \cevns recoil energy and scattering angle, thereby allowing event-by-event reconstruction of the neutrino energy. We estimate that a 60:40 He:CF$_4$ gas mixture at atmospheric pressure offers a good trade-off between total target mass and good directionality and could deliver a detection of the angular distribution of \cevns, even under pessimistic background conditions. We project the sensitivity of 1 and 10 m$^3$-scale detectors in the context of several physics cases, including: the measurement of the Standard Model \cevns cross section, reconstruction of the flavour-dependent neutrino fluxes, observing the neutrino-induced Migdal effect, constraints on beyond-Standard Model neutrino interactions, and probing 10-eV-scale sterile neutrinos.
\end{abstract}

\maketitle

\section{Introduction}
When it was first proposed in 1974~\cite{Freedman:1973yd}, Freedman suggested that it would be an ``act of hubris'' to imagine that detecting the coherent elastic scattering of neutrinos on nuclei (\cevns) would be experimentally feasible. Despite the \cevns cross section being large by the standards of neutrino interactions, the resulting nuclear recoil energies generated by available neutrino sources were too small for the technology available at the time. However, in 2017, the first measurement of \cevns was finally achieved by the COHERENT collaboration~\cite{COHERENT:2022nrm}, who employed a 14.6~kg CsI scintillator exposed to the stopped-pion neutrino flux generated by the Spallation Neutron Source (SNS) at Oak Ridge National Laboratory~\cite{Akimov:2017ade}. The $\lesssim50$~MeV neutrinos created by the pion decays at rest generate tens of keV nuclear recoils through \cevns, which are measurable by modern detectors.

\cevns has since transitioned from discovery to precision science~\cite{Abdullah:2022zue}. Several other measurements have been made by the COHERENT collaboration~\cite{COHERENT:2022nrm,Adhikari:2026qrv} using liquid argon~\cite{COHERENT:2020iec, COHERENT:2020ybo} and germanium~\cite{COHERENT:2024axu} detectors.
\cevns has also now been detected using two other important sources of neutrinos, namely solar neutrinos by LZ~\cite{Akerib:2025xla}, XENONnT~\cite{XENON:2024ijk} and PandaX~\cite{PandaX:2024muv}, and reactor neutrinos by CONUS+~\cite{Ackermann:2025obx}. Taken together, these measurements have inspired the neutrino community to take advantage of the unique ability of \cevns to test beyond-Standard-Model neutrino interactions (see, for example, Refs.~\cite{Cerdeno:2016sfi, Dent:2016wcr, Dutta:2017nht, Dent:2017mpr, Abdullah:2018ykz, Denton:2018xmq, Amaral:2020tga, Miranda:2020zji,delaVega:2021wpx, Majumdar:2021vdw, Li:2022jfl, Schwemberger:2022fjl, AtzoriCorona:2022moj, AtzoriCorona:2022jeb, Amaral:2023tbs, Giunti:2023yha, AristizabalSierra:2024nwf, DeRomeri:2024iaw, DeRomeri:2024hvc, DeRomeri:2025csu, Maity:2024aji, Blanco-Mas:2024ale, Xia:2024ytb}) that are more difficult to access using e.g.~neutrino-electron scattering, inverse beta-decay, or neutrino oscillation experiments.

\cevns proceeds through the exchange of a $Z$ boson and is flavour blind at tree level~\cite{Sehgal:1985iu,Tomalak:2020zfh,Mishra:2023jlq}. As the name implies, the cross section is coherently enhanced by the entire nucleus, up to small incoherent corrections from the nuclear structure that become important towards high momentum transfer~\cite{Hoferichter:2020osn, Payne:2019wvy, AbdelKhaleq:2024hir, VanDessel:2020epd}. The cross section scales approximately with the square of the number of neutrons in the nucleus, which means that most measurements have sought to take advantage of the higher statistics that come with using heavier target nuclei. In addition, the only experimental signatures of \cevns exploited at present involve measuring the deposition of nuclear recoil energy into a cumulative heat, scintillation, or ionisation signal.

Although \cevns is being pursued by several existing and near-future detectors, the current experimental landscape presents an opportunity for a different type of measurement that could complement the global \cevns program. We propose here the use of a 1--10~m$^3$-scale gaseous time-projection chamber (TPC) instrumented with a highly segmented charge-amplification and readout plane. The primary aim of this experiment would be to perform \textit{direction-sensitive} measurements of \cevns by three-dimensionally imaging the mm--cm-length recoil tracks deposited by \cevns-induced recoils of the gas nuclei.

Highly-segmented micro-pattern gas detectors (MPGDs) that can achieve sub-mm spatial reconstruction of recoils in three dimensions are already a reality---see, for example, summaries of this concept presented in~\cite{Vahsen:2020pzb, Vahsen:2021gnb, OHare:2022jnx, Surrow:2022ptn}, which we refer to as ``recoil imaging''. In the context of rare-event searches, recoil imaging has seen significant advances over the last few decades, motivated in part by the goal of finding a new approach for dark matter direct detection that is able to circumvent~\cite{Grothaus:2014hja, OHare:2015utx, Mayet:2016zxu, OHare:2017rag, OHare:2020lva} the so-called ``neutrino fog'' limit imposed upon conventional recoil-based experiments~\cite{Billard:2013qya,Ruppin:2014bra,OHare:2016pjy, Dent:2016iht, Dent:2016wor, Gelmini:2018ogy,OHare:2021utq,Akerib:2022ort,Aalbers:2022dzr,Carew:2023qrj,Maity:2024vkj,XLZD:2024nsu,Dent:2025drd}. So the experiment we propose here has natural synergies with the separate vision of a large-scale ``recoil observatory'' to be ultimately installed in a deep underground laboratory to perform directional measurements of astrophysical neutrinos~\cite{Lisotti:2024fco,Shekar:2025xhx} and dark matter~\cite{Mayet:2016zxu}; as is the goal of the \Cygnus Consortium~\cite{Vahsen:2020pzb,Schueler:2022lvr,Ghrear:2024rku,Battat:2016xxe,Ikeda:2021ckk} and CYGNO collaboration~\cite{Baracchini:2020btb,Amaro:2023dxb,Almeida:2023tgn,CYGNO:2023gud,Amaro:2025pms}.

The first exploration of the idea of directional \cevns measurements at a neutrino source was detailed in Ref.~\cite{Abdullah:2020iiv}. The goal of the present study is to expand on this physics case and determine the optimal set of detector parameters to perform the most scientifically interesting measurement of \cevns at SNS. These parameters will include the volume, gas mixture and gas pressure, for which there is a trade-off between the total event rate and the directional performance. We will use these parameters to forecast sensitivity to a range of possible science cases, including pure-Standard Model (SM) measurements such as the reconstruction of the \cevns cross section or the incoming neutrino fluxes, as well as beyond-Standard Model (BSM) interactions and sterile neutrinos.

The rest of the paper is laid out as follows. We begin in Sec.~\ref{sec:eventrate} by calculating the event rate of \cevns on a fiducial set of target nuclei under the Standard Model. Then in Sec.~\ref {sec:detector} we describe the basic detector design and discuss several important performance limitations such as the finite energy and angular reconstruction. Then in Sec.~\ref{sec:SM} we forecast sensitivity estimates for measurements in the context of the Standard Model, before discussing BSM measurements in Sec.~\ref{sec:BSM}. We conclude in Sec.~\ref{sec:conc}

\section{CEvNS event rate}\label{sec:eventrate}
We begin by deriving the expected rate of \cevns events within the Standard Model, given a fiducial model for our experimental setup, namely a gas TPC containing a 60:40 ratio \hecf gas mixture. This fiducial setup, in terms of the gas mixture and volume, is motivated in Sec.~\ref{sec:detector}.

\subsection{Neutrino fluxes}\label{sec:fluxes}
\begin{figure}
    \centering
    \includegraphics[width=0.99\linewidth]{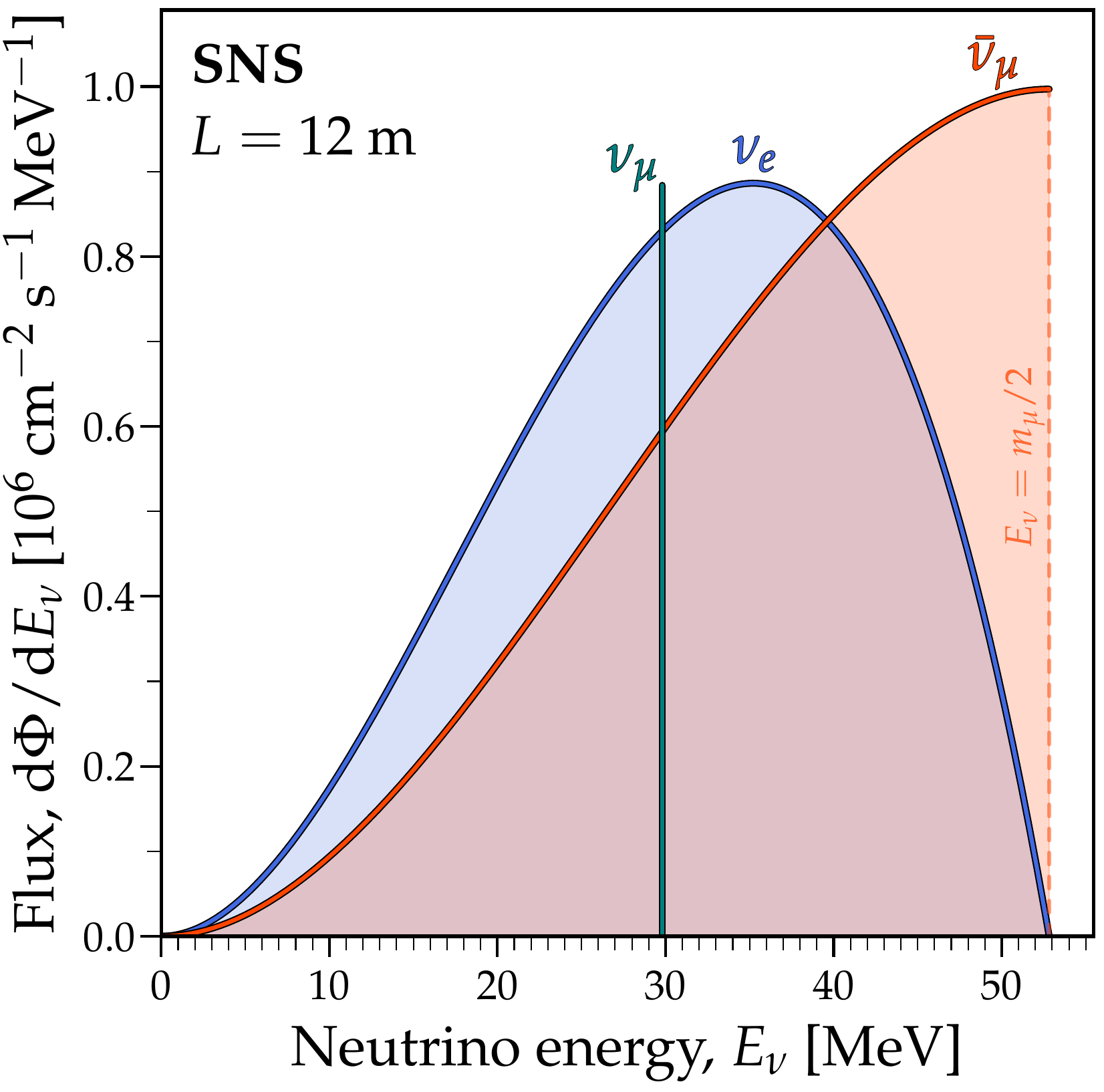}
    \caption{The SNS neutrino flux model adopted in this work, which consists of two continuous fluxes of delayed $\nu_e$ and $\bar{\nu}_\mu$ as well as the mono-energetic line at $E_\nu \approx 30$~MeV corresponding to the prompt flux of $\nu_\mu$. The assumed effective source-to-detector distance is 12 metres.}
    \label{fig:NuFlux}
\end{figure}
The SNS is designed to generate pulsed neutron beams by colliding $\sim1$~GeV protons with a flowing liquid-mercury target. Since these collisions also produce charged pions, which quickly stop in the target and decay at rest, the SNS also generates several pulsed fluxes of neutrinos as a by-product. There are three isotropic fluxes of neutrinos produced in these collisions: a monochromatic line of $\nu_\mu$ at \mbox{$E_v=\left(m_\pi^2-m_\mu^2\right) / 2 m_\pi \simeq 30~ \mathrm{MeV}$} from the prompt pion decays, and two continuous fluxes of $\nu_e$ and $\bar{\nu}_\mu$ from delayed $\mu^+$ decays which have a lifetime of around 2.2 $\upmu$s~\cite{COHERENT:2021yvp}. The spectra of the three fluxes are well-described by the following analytic functions,
\begin{equation}
    \begin{aligned}
& \mathcal{F}_{\nu_\mu}\left(E_\nu\right)=\frac{2 m_\pi}{m_\pi^2-m_\mu^2} \delta\left(1-\frac{2 E_\nu m_\pi}{m_\pi^2-m_\mu^2}\right), \\
& \mathcal{F}_{\nu_e}\left(E_\nu\right)=\frac{192}{m_\mu}\left(\frac{E_\nu}{m_\mu}\right)^2\left(\frac{1}{2}-\frac{E_\nu}{m_\mu}\right), \\
& \mathcal{F}_{\bar{\nu}_\mu}\left(E_v\right)=\frac{64}{m_\mu}\left(\frac{E_\nu}{m_\mu}\right)^2\left(\frac{3}{4}-\frac{E_\nu}{m_\mu}\right) .
\end{aligned}
\end{equation}
For the muon and pion masses we take \mbox{$m_\mu = 105.65$~MeV} and \mbox{$m_\pi = 139.57$~MeV}~\cite{ParticleDataGroup:2024cfk} respectively. We apply a normalisation constant to these functions to calculate the time-averaged total flux, set by the number of proton collisions per unit time and the number of neutrinos produced per collision, i.e.~
\begin{equation}
    \frac{\drm \Phi_{\nu_i}}{\drm E_\nu} = \frac{r n_{\rm POT}}{4\pi L^2} \mathcal{F}_{\nu_i}(E_\nu) \, .
\end{equation}
We fix $r = 0.087$ to be the number of pions produced per proton-mercury collision, and \mbox{$n_{\rm POT} = 1.728\times 10^{23}~{\rm year}^{-1}$} as the number of protons-on-target. COHERENT has put forward a plan to install a heavy-water detector to provide a $<$~5\% measurement of the total neutrino flux normalisation in the near future~\cite{COHERENT:2021xhx}. Throughout, we will assume $L = 12$~m as the effective distance from the neutrino source to the detector, based on estimates for space availability at the site. 
The three neutrino fluxes are visualised in Fig.~\ref{fig:NuFlux}.

\subsection{Coherent elastic neutrino-nucleus scattering}\label{sec:cevns}
Given a nucleus $N$ of mass $m_N$, the \cevns process $\nu N \rightarrow \nu N$ occurs through neutral current interactions via $Z$ exchange in the SM. The tree-level cross-section is given by,
\begin{equation}
   \frac{\drm \sigma}{\drm E_r} = \frac{G_F^2 m_N F^2(q) g_V^2}{\pi} \left( 1 - \frac{m_N E_r}{2E_\nu^2} \right) \, ,
    \label{eq: SM CEVNS CS}
\end{equation}
where $G_F = 2.3016\times 10^{-22}$~cm$\,$MeV$^{-1}$ is the Fermi constant. For a given nucleus with atomic number $Z$, mass number $A$ and the number of neutrons $N=A-Z$, the SM vector coupling of the nucleus is $g_V = Z g_p^V + N g_n^V$. Here, $g_p^V = 2 g_u^V + g_d^V = \frac{1}{2} - 2\sin^2{\theta_w}$ and $g_n^V = g_u^V + 2g_d^V = -\frac{1}{2}$, which signify the couplings with protons and neutrons respectively, in terms of the $u$ and $d$ quark couplings. We use the value $\sin^2{\theta_w} = 0.2387$ for the weak mixing angle under the $\overline{\rm MS}$ renormalisation scheme~\cite{ParticleDataGroup:2024cfk}. The form factor, $F(q)$, is used to capture the effects of nuclear structure, which become important at large momentum transfer, $q=\sqrt{2m_N E_r}$. We adopt the commonly-used Helm ansatz for the form factor~\cite{Lewin:1995rx}:
\begin{equation}
    F(q) = \frac{3j_1(q r_n)}{q r_n} \exp{\left(\frac{-q^2s^2}{2}\right)} \, ,
    \label{eq: Form Factor}
\end{equation}
where $j_1(q)$ is the spherical Bessel function of the first kind. The nuclear radius is parameterised by \mbox{$r_n = \sqrt{\frac{5}{3}\left(R_{\rm min}^2 - 3s^2\right)}$}, where $s=0.9$~fm is a measure of the nuclear skin thickness and $R_{\rm min}$ is the root-mean-square radius of the proton distribution in the nucleus. We use $R_{\rm min} = (1.6755,\,2.4702,\,2.8976)$~fm~\cite{Angeli:2013epw} for the three nuclei we study here: $(^4{\rm He},\,^{12}{\rm C},\,^{19}{\rm F})$. The form factor is a very minor correction to the \cevns rate for these light nuclei, so we will not consider nuclear structure uncertainties in this study.

Typically, \cevns is measured by detecting the fraction of the nuclear recoil energy $E_r$ that is converted into, e.g.~scintillation, heat or ionisation. This means that, given knowledge of details such as the detector efficiency and quenching etc., the measurable signal exploited by \cevns experiments present is ultimately linked to the distribution of recoil energies, $\textrm{d}R/\textrm{d}E_r$. For this proposal, we will expand the space of observables to also include the recoil direction $\mathbf{\hat{q}}_r$, in addition to the recoil energy. 

\begin{figure*}
    \centering
    \includegraphics[height=0.4765\linewidth]{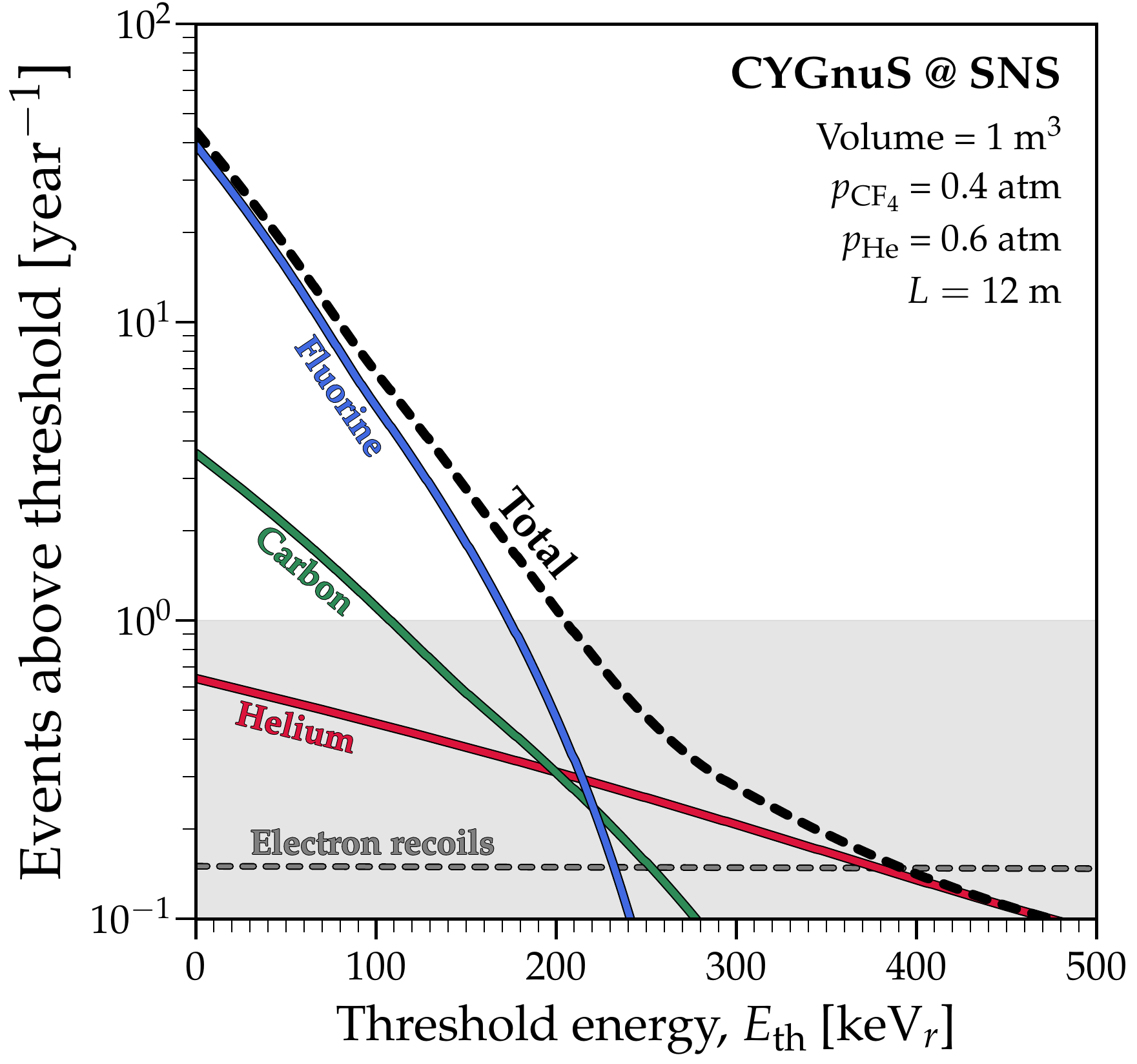}
        \includegraphics[height=0.47\linewidth]{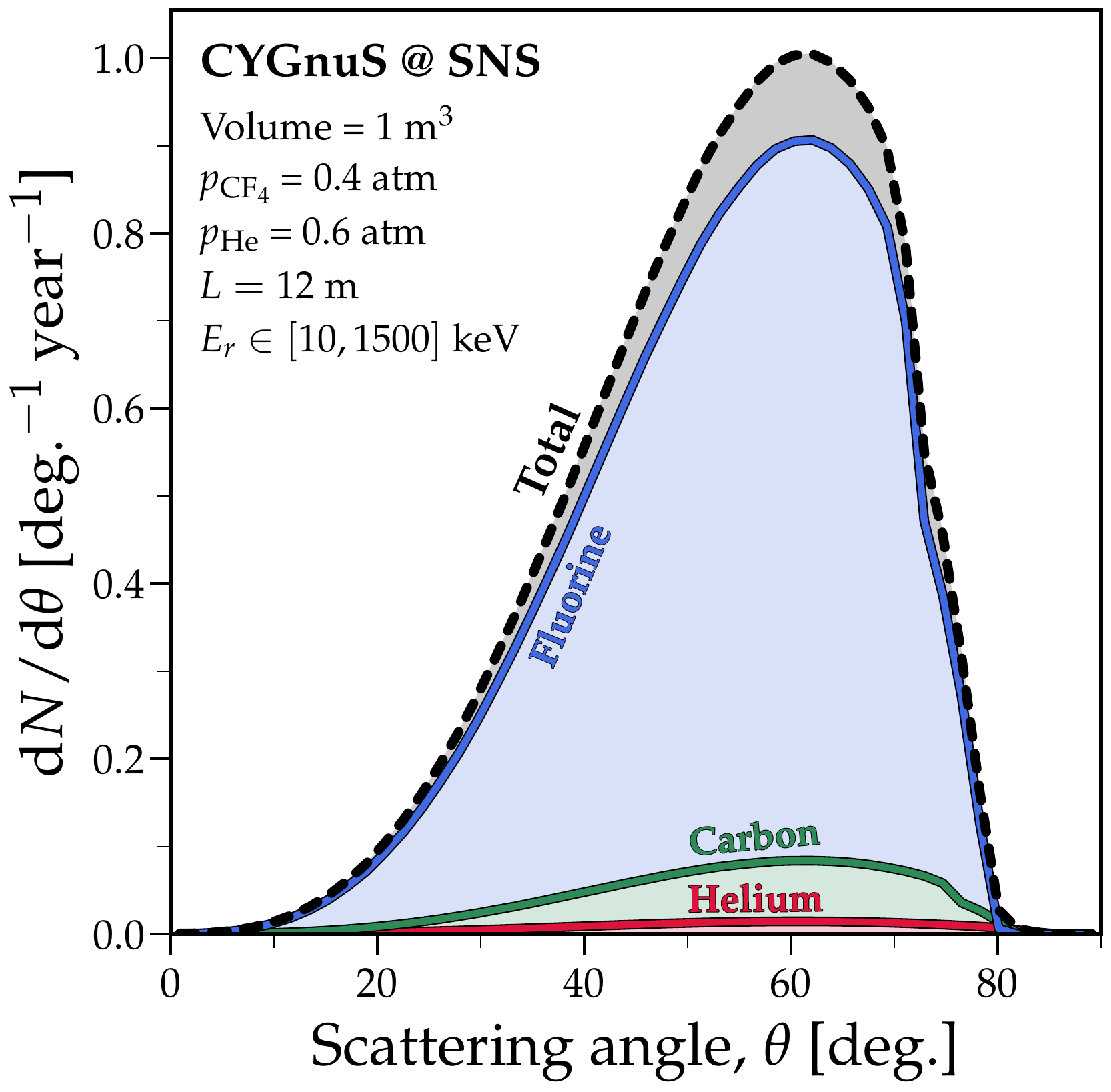}
    \caption{{\bf Left}: Expected number of events per year in a $V = 1$~m$^3$ experiment as a function of the recoil energy threshold $E_{\rm th}$, defined by integrating Eq.~(\ref{eq:d2RdEdOmega}) over all angles and for energies $E_r>E_{\rm th}$. {\bf Right}: Distribution of events in the same gas mixture as the left-hand panel, as a function of the scattering angle, $\theta$. This is calculated by integrating Eq.~(\ref{eq:d2RdEdOmega}) over energies in the range $E_r \in [10,1500]$~keV, and then changing variables to $\theta$ as opposed to $\cos{\theta}$. In both panels, as well as the total yearly event rate (black dashed line), we also show the number of events attributed to each nucleus in our fiducial 60:40 \hecf gas mixture.}
    \label{fig:EventRate}
\end{figure*}

To find the combined distribution of recoil energies and angles, we compute the double-differential event rate per unit detector mass, $\drm^2 R/\drm E_r \drm \Omega_r$, where $\drm\Omega_r$ is a solid angle element around $\mathbf{\hat{q}}_r$. The expression for this is the following,
\begin{equation}
    \frac{\drm^2R_{\nu_i, T_j}}{\drm E_r\drm\Omega_r} = N_{T_j} \int \drm E_\nu \drm\Omega_\nu \frac{\drm^2\Phi_{\nu_i}}{\drm E_\nu \drm \Omega_\nu} \frac{\drm^2\sigma_{\nu_i,T_j}}{\drm E_r \drm\Omega_r} 
    \label{eq: DRS} \, ,
\end{equation}
where $\nu_i$ labels one of the three neutrino flavours and $T_j$ labels one of the nuclear species present in the target, e.g.~$T_j = (^4{\rm He},\,^{12}{\rm C},\,^{19}{\rm F})$, and $N_{T_j}$ is the number of target nuclei per unit detector mass. Since $\sigma$ is flavour-independent in the SM (up to radiative corrections), the label $\nu_i$ will only be relevant when we include BSM contributions. The first quantity inside the integral is the double-differential neutrino flux, which in the case of a point source of neutrinos can be written straightforwardly as,
\begin{equation}
    \frac{\drm^2\Phi_{\nu_i}}{\drm E_\nu \drm\Omega_\nu} = \frac{\drm\Phi_{\nu_i}}{\drm E_\nu} \delta \left(\mathbf{\hat{q}}_\nu-\mathbf{\hat{q}}_0 \right) \, ,
    \label{eq: 2-D Flux}
\end{equation}
where $\mathbf{\hat{q}}_{\nu}$ is the incoming neutrino direction and $\mathbf{\hat{q}}_0$ is the unit vector pointing from the source to the detector. 

To relate the double-differential cross section, $\textrm{d}^2\sigma/\textrm{d}E_r \textrm{d}\Omega_r$, to Eq.~(\ref{eq: SM CEVNS CS}), we assume the scattering is azimuthally symmetric about $\mathbf{\hat{q}}_0$. This means our solid angle element can be written as $\drm\Omega_r = 2\pi \drm\cos\theta$ if we work, as we will here, in a coordinate system in which $\theta$ is measured relative to $\mathbf{\hat{q}}_0$. Enforcing conservation of momentum in the relativistic scattering kinematics, one can derive the following expression for the scattering angle in terms of the neutrino energy and the energy of the recoil,
\begin{equation}\label{eq:costh}
    \cos\theta = \mathbf{\hat{q}}_\nu \cdot \mathbf{\hat{q}}_r = \frac{E_\nu + m_N}{E_\nu} \sqrt{\frac{E_r}{E_r + 2m_N}} \, ,
\end{equation}
Performing the integral above, we arrive at~\cite{OHare:2015utx},
\begin{equation}\label{eq:d2RdEdOmega}
    \frac{\drm^2R_{\nu_i,T_j}}{\drm E_r \drm\Omega_r} = \frac{N_T}{2\pi} \frac{\mathcal{E}^2}{y} \left. \left(\frac{\drm\sigma_{\nu_i,T_j}}{\drm E_r} \frac{\drm\Phi_{\nu_i}}{ \drm E_\nu} \right) \right| _{E_\nu = \mathcal{E}} \, .
\end{equation}
where, 
\begin{equation}
   y=\sqrt{\frac{m_N^2E_r}{E_r + 2m_N}}, \quad  \mathcal{E} = \left(\frac{\mathbf{\hat{q}}_r \cdot  \mathbf{\hat{q}}_0}{y} - \frac{1}{m_N}  \, \right)^{-1}.
\end{equation}
The smallest neutrino energy that can generate a recoil of energy $E_r$ is $E_\nu^{\rm min} = \sqrt{m_N E_r/2}$. In the non-directional calculation of the rate (i.e.~$\textrm{d}R/\textrm{d}E_r$), this lower limit to the integral must be enforced. However, for the directional calculation $E_\nu^{\rm min}$ corresponds to $\cos{\theta} = 0$, and so this cut-off is already taken into account by not computing the rate for unphysical (negative) values of $\cos\theta$.

We will evaluate our physics sensitivities based on the distribution $\drm^2R_{\nu_i, T_j}/\drm E_r \drm\Omega_r$ summed across the three target nuclei and three fluxes. We do \textit{not} expect our envisioned detector to be able to resolve the event timing precisely enough, so we discount the possibility of including this information as a third observable, and assume all three fluxes of neutrinos are detected simultaneously. The arrival-time profile of events within each beam pulse would provide further information about the neutrino flavour, but this requires sub-microsecond timing resolution, which is not accessible in a gas TPC that exclusively measures drifted ionisation. That said, timing information is still available in general and would be used to measure the background between beam pulses.

\subsection{Event rate}\label{sec:eventrate_plots}
Let us now fix a fiducial target mixture of He and \cffour. A volume of 1~m$^3$ and a 60:40 ratio of \hecf at 1~atmosphere (101325 Pa) of pressure and a temperature of 293~K, gives a total mass in (He, C, F) of $M_{T_j} = (0.1, 0.2, 1.26)$~kg. For a threshold energy of $E_{\rm th} = 10$~keV, we calculate an expected $(0.62,\, 3.27,\, 32.97)$ \cevns events for each nucleus for that gas mixture and volume. The gas ratio is a flexible parameter in the design of this experiment, but for concreteness we will fix it at this ratio for most of the discussion and explain the experimental motivation behind this choice more carefully in Sec.~\ref{sec:detector}.

In Fig.~\ref{fig:EventRate} we show the expected number of events in a 1~m$^3$ detector using a 60:40 \hecf gas mixture (left-hand panel), as well as the expected distribution of lab-frame neutrino-nucleus scattering angles, $\theta$ (right-hand panel). Reiterating what was stated above, the left-hand plot shows that we expect around 37 events per year if we set a conservative $\sim10$~keV recoil energy threshold. The majority of these events would be fluorine recoils, with only a handful expected due to scattering on the carbon nucleus in \cffour---this is a result of the $\sim N^2$ dependence of the \cevns cross section. We would require volumes larger than 1~m$^3$ (or a much more helium-rich gas mixture) to observe helium recoils; however, these recoils would have higher energies and so would generate longer tracks, making their track reconstruction superior to that of fluorine and carbon recoils. 

As shown in the right-hand of Fig.~\ref{fig:EventRate}, we expect recoils to emerge with a most probable scattering angle around 60 degrees. The sharp cut-off in the distribution at $\sim 80^\circ$ is set by the assumed lower energy threshold of 10~keV. Due to the azimuthal symmetry around the source direction, we therefore expect the true \cevns recoil track directions in three dimensions to point towards an annulus with inner/outer radius $\sim 10$ and $80^\circ$ (although the limited angular reconstruction will blur this signal in practice, as discussed below). The slight discontinuity in the angular distribution between 70 and 80$^\circ$ is due to the $\nu_\mu$ line. The standard deviation of this angular distribution is between 13--15$^\circ$ across all target nuclei and fluxes---we will use this angular spread as a point of comparison when we discuss the angular resolution achievable in this gas mixture in Sec.~\ref{sec:performance}.

\section{Detector model}\label{sec:detector}
\begin{figure}
    \centering
    \includegraphics[width=0.99\linewidth]{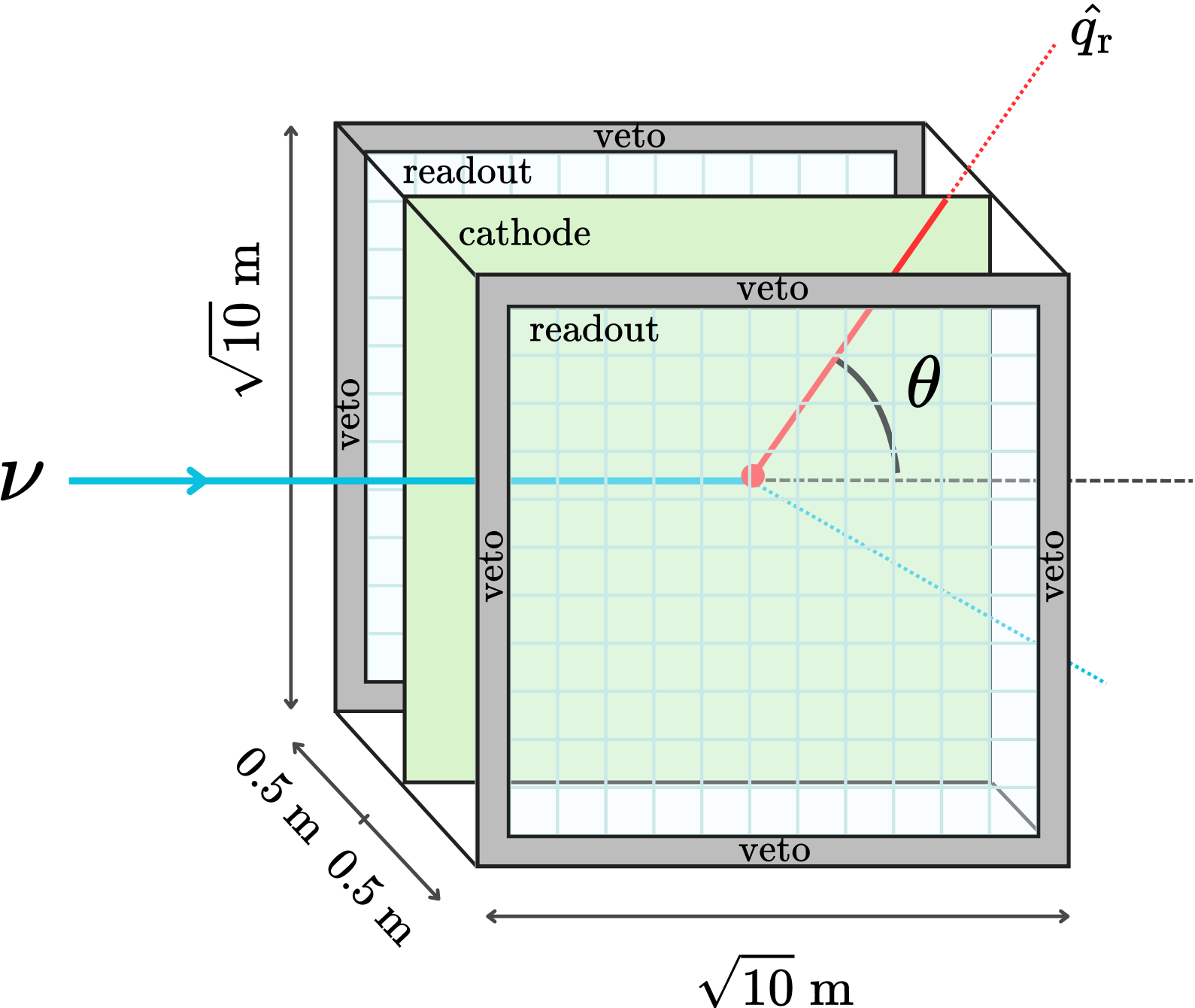}
    \caption{Diagram of the anticipated dimensions of a 10~m$^3$ directional gas TPC. The drift axis would be kept short, limiting the maximum drift distance to 50 cm so as to reduce the impact of diffusion, as this has the largest impact in washing out the directionality of the charge track (discussed further in Fig.~\ref{fig:angular_resolution}). An initial 1~m$^3$ version would have the same design but with the readout planes scaled down.}
    \label{fig:detector}
\end{figure}
Having established the expected event numbers, energies, and directions of \cevns recoils for a \hecf gas mixture, we will now motivate our envisioned experimental setup and discuss the parameterisation of its performance.

A simplified conceptual design of the detector is shown in Fig.~\ref{fig:detector}. It will consist of a TPC filled with a \cffour-based gas mixture where ionisation electrons produced by recoil tracks are drifted by an applied electric field and then amplified and detected at a two-dimensional readout plane utilising micro-patterned gaseous detector (MPGD) technologies.

The final gas may be pure \cffour at atmospheric or lower pressure. This is the option we have parameterised carefully in our detector performance model. The addition of helium gas would be used to ensure the total pressure remains atmospheric, and since helium is a very light nucleus, it will not severely impact the performance of the detector. A gas target at atmospheric pressure would only require a relatively inexpensive gas-tight vessel to contain the target gas, rather than a true vacuum vessel capable of withstanding order 1 atm pressure differences. The addition of helium and atmospheric pressure operation could result in significant cost savings during construction. On the flip side, a proper vacuum vessel would provide greater flexibility during operation. It would enable operation with sub-atmospheric total gas pressure, and would furthermore have the advantage of allowing a full evacuation before filling the detector with the final target gas. This means sufficiently high gas purity could be reached more rapidly.

When one of the nuclei present in the target gas recoils with an energy $\gtrsim$keV it will leave a mm- to cm-scale track of ionisation, which then drifts along the $z$-axis, due to a uniform electric field applied across the gas volume, towards a readout plane. During this drift, the ionised electrons repeatedly scatter off gas molecules and re-accelerate in the electric field. The resulting effective motion is that the ionisation travels with a uniform effective velocity in the direction of the E-field, in addition to undergoing three-dimensional diffusion which smears out the shape of the initial charge cloud (see e.g.~Ref.~\cite{Vahsen:2021gnb} for more discussion and visualisations of this).

When the drifted electrons arrive at the readout plane, they first pass through an MPGD-based gas amplification device (we foresee a Micromegas), before the amplified charge induces a signal in two stacked, highly-segmented anode planes. One plane is segmented with strips in $x$, the other has strips in $y$. As a result, the $x$-$y$ position of the ionisation is detected as it arrives. The readout strips will have a pitch at the 100~$\upmu$m scale, sufficiently small so that the aforementioned diffusion will dominate the position resolution. The third coordinate $z$ (along the drift axis), is reconstructed using the arrival time of charges on the strips, assuming knowledge of the drift velocity of electrons through the gas mixture. 

Considerations of the straggling of the nuclear recoil, the effects of charge diffusion during the drift, as well as the limitations in resolution set by the finite number of ionised electrons, will all be incorporated into our model for the finite energy and angular resolution discussed below.

For an initial experiment large enough to make the first measurement of the directionality of \cevns, we propose a $\sim$1~m$^3$ gas vessel. We will assume up to three years of real-time operation---note that we already account for the duty cycle of the SNS beam in the definition of $n_{\rm POT}$. To explore the further physics potential of a larger-scale experiment, we will also show results for a $10~{\rm m}^3 \times 3 \,{\rm years}$ exposure, which is also feasible given the space availability.

Our chosen source-to-detector distance of 12~m is far enough that we can reasonably approximate the neutrinos as originating from a point source. We will assume that not only will the recoil track directions be reconstructed in three dimensions, but so will the \textit{positions} of the tracks within the detector volume. This latter assumption is required if we are to assume the $\cos\theta$ angle can be defined for each recoil as one of the primary observables. This is achieved in practice through two different ways. The absolute $x$-$y$ track position is reconstructed to sub-cm precision because the readout plane is highly segmented. The absolute track position along the TPC drift direction, $z$, requires some knowledge of the absolute time taken for the charge to drift from the point of energy deposition to the readout plane. For neutrino events, this time may be inferred from the delay between the most recent SNS pulse and the TPC event time. For non-prompt events, similar knowledge can be obtained from the amount of transverse diffusion of the charge cloud~\cite{Lewis:2014poa} as this also depends on the drift time.\footnote{An alternative method is also possible in negative ion drift gases like SF$_6$ thanks to the presence of minority carriers~\cite{Snowden-Ifft:2014taa, Phan:2016veo} (ion species that drift at different velocities). This is not available in pure \cffour, which is an electron-drift gas.} 


In most designs for gas-based recoil imaging experiments, the optimal dimensions for the volume would likely be one with an uneven aspect ratio in which the two dimensions along the readout plane would be longer than the drift axis dimension. It is important to limit the maximum distance a charge track must drift because, as we will show, the deleterious impact of diffusion will be the major limiting factor in achieving good directional reconstruction of low-energy tracks.

As shown Fig.~\ref{fig:detector} a 10~m$^3$ volume detector could take on a dual-readout $1\,{\rm m}\times \sqrt{10}\,{\rm m}\times \sqrt{10}\,{\rm m}$ design. Using a back-to-back TPC design like this (i.e.~two readouts on either side of the volume with the central cathode running through the middle) would ensure the maximum possible drift distance is kept to only 50 cm.

The detector scenario presented here should be considered a conceptual design, and ignores complications beyond those needed to explore the physics potential and sensitivity. For example, a true detector based on strips would detect signals using attached charge-sensitive preamplifiers. This implies a maximum length for strips, to keep their capacitance and the resulting pre-amplifier noise floor acceptable. This in turn means some further segmentation of the $x-y$ readout planes is required. In addition, the pre-amplifier's shaping characteristics affect the reconstruction performance, and the detector needs to incorporate cooling for the preamplifiers. We ignore all such complications here. Finally, all dimensions stated here, such as drift length and readout plane pitch must be carefully optimised when planning the real experiment---the values we choose here are indicative only. Such optimisation studies are in progress. A first optimisation of the strip and amplification region dimensions of a Micromegas with $x$--$y$ strip readout can be found in Ref.~\cite{Ghrear:2024bxi}.

\subsection{Performance parameterisation}\label{sec:performance}
The two key observables that we will base our analysis on are the nuclear recoil energy $E_r$ and the track angle with respect to the neutrino source direction, $\theta$. To optimise our planned detector, we construct a parameterised model that predicts how accurately these two quantities can be measured in terms of an energy resolution, $\sigma_{E}$, and an angular resolution, $\sigma_{\theta}$. 

These two resolutions both depend on the recoil energy, because this determines both the amount of ionisation available to be measured and the length of the recoil track. The angular resolution is also strongly affected by the gas pressure, which modifies both the initial track length and the diffusion of ionised electrons as they drift through the gas volume. For a track axis to be detectable, this initial track must be longer than the diffusion scale of electrons in the detector, hence the performance will degrade at low energies and high gas pressures when the tracks lengths become very short.

The energy- and pressure-dependence of the angular resolution, which we will write as $\sigma_\theta(p_{\rm CF_4},E_r)$, is the quantity that most substantially impacts the ability to reconstruct the angular distribution of \cevns recoils. Since the standard deviation of the true underlying recoil distribution is around $13$--$15^\circ$ (see Fig.~\ref{fig:RecoilDistributions}), heuristically we wish to aim for an axial angular resolution of $\sigma_\theta \lesssim 30^\circ$ across the majority of the observable recoil spectrum so as to prevent the limitations in the track axis reconstruction from overwhelming the underlying angular distribution.

However, there is a trade-off between event statistics and obtaining good directional performance: At higher gas pressures, we expect more recoil events, but the angular resolution will be worse. In what follows, we detail our model for the energy- and pressure-dependent detector performance, which we use to identify the gas pressure that is a good compromise between recoil event statistics and detector resolution.

We assume for simplicity that all our nuclear recoil energies are quoted as true recoil energies, or in other words, that they have already been corrected for quenching. Similarly, we will assume a flat event-level efficiency curve equal to unity above the detector's energy threshold. Both of these simplifications are factored into the conservative choice for our (true) nuclear recoil energy threshold of $E_{\rm th} = 10$~keV$_r$, which we take to be the energy above which those corrections do not severely impact the rate. It is still possible to measure events at lower energies, but their reconstruction, particularly of their directions, will be poor.



\subsection{Energy resolution}
We assume that the reconstructed value of the recoil energy follows a Gaussian distribution centred on the true recoil energy with an energy-dependent spread $\sigma_E(E_r)$. In practice this means that, for a given true underlying value of $E_r$ sampled from $\textrm{d}^2R/\textrm{d}E_r\textrm{d}\Omega_r$, we draw a `measured' recoil energy $E^\prime_r$ from the distribution,
\begin{equation}
    K_E\left(E_{r}, E_{r}^{\prime}\right)=\frac{1}{\sqrt{2 \pi} \sigma_E(E_{r})} \exp\left(-\frac{\left(E_{r}-E_{r}^{\prime}\right)^{2}}{2 \sigma_{E}^{2}(E_{r})}\right) \, .
\end{equation}
We use the following simple function for the fractional energy resolution, 
\begin{equation}\label{eq:energyresolution}
    \frac{\sigma_{E}}{E_r}=\sqrt{(0.02)^2 + (0.1)^2\left(\frac{5.9\,{\rm keV}_{\rm ee}}{E_rQ(E_r)}\right)} \, ,
\end{equation}
where $Q(E_r)$ is the nuclear-recoil ionisation quenching factor, which is modelled using the Lindhard theory of energy partition in ion--atom collisions~\cite{Lindhard1963}. 
In this framework, the fraction of recoil energy transferred to electronic channels (ionisation and excitation) is written as
\begin{equation}\label{eq:quenching}
Q(E_r)=\frac{k\,g(\epsilon)}{1+k\,g(\epsilon)}\,,
\end{equation}
where $\epsilon = 11.5\,E_r\,Z^{-7/3}$ is the dimensionless recoil energy and $g(\epsilon)=3\epsilon^{0.15}+0.7\epsilon^{0.6}+\epsilon$. 
We adopt \mbox{$k=0.133\,Z^{2/3}A^{-1/2}$}, following the commonly used analytic approximation to the Lindhard solution~\cite{LewinSmith1996}. 
This semi-empirical parameterisation provides an adequate description of the energy partition between nuclear and electronic stopping at keV-scale recoil energies and is widely employed to estimate ionisation yields in rare-event detectors. For the simulated fluorine recoils used in our analysis, discussed further below, Eq.~(\ref{eq:quenching}) with $Z=9$ and $A=19$ matches the simulation to within 13\%.

Equation~(\ref{eq:energyresolution}) reproduces  typical experimental results for TPCs with MPGD-based charge readout, which tend to achieve a fractional energy resolution of $\sigma_E/E_r$ of order 10--20\% around the 5.9 keV energy of the $^{55}$Fe source commonly used for calibration. The $(E_r Q)^{-1/2}$ dependence reflects the typical dependence of fluctuations on the mean visible ionisation, while the floor in the resolution function of 2\% is simply enforced to avoid the energy reconstruction becoming arbitrarily good at high energies. This latter addition has a minimal impact on our results, as the performance is primarily driven by the low-energy dependence between 10--50~keV$_r$ due to the exponentially falling event rate.

\subsection{Angular resolution}
\begin{figure}
    \centering
  \includegraphics[width=0.99\linewidth]{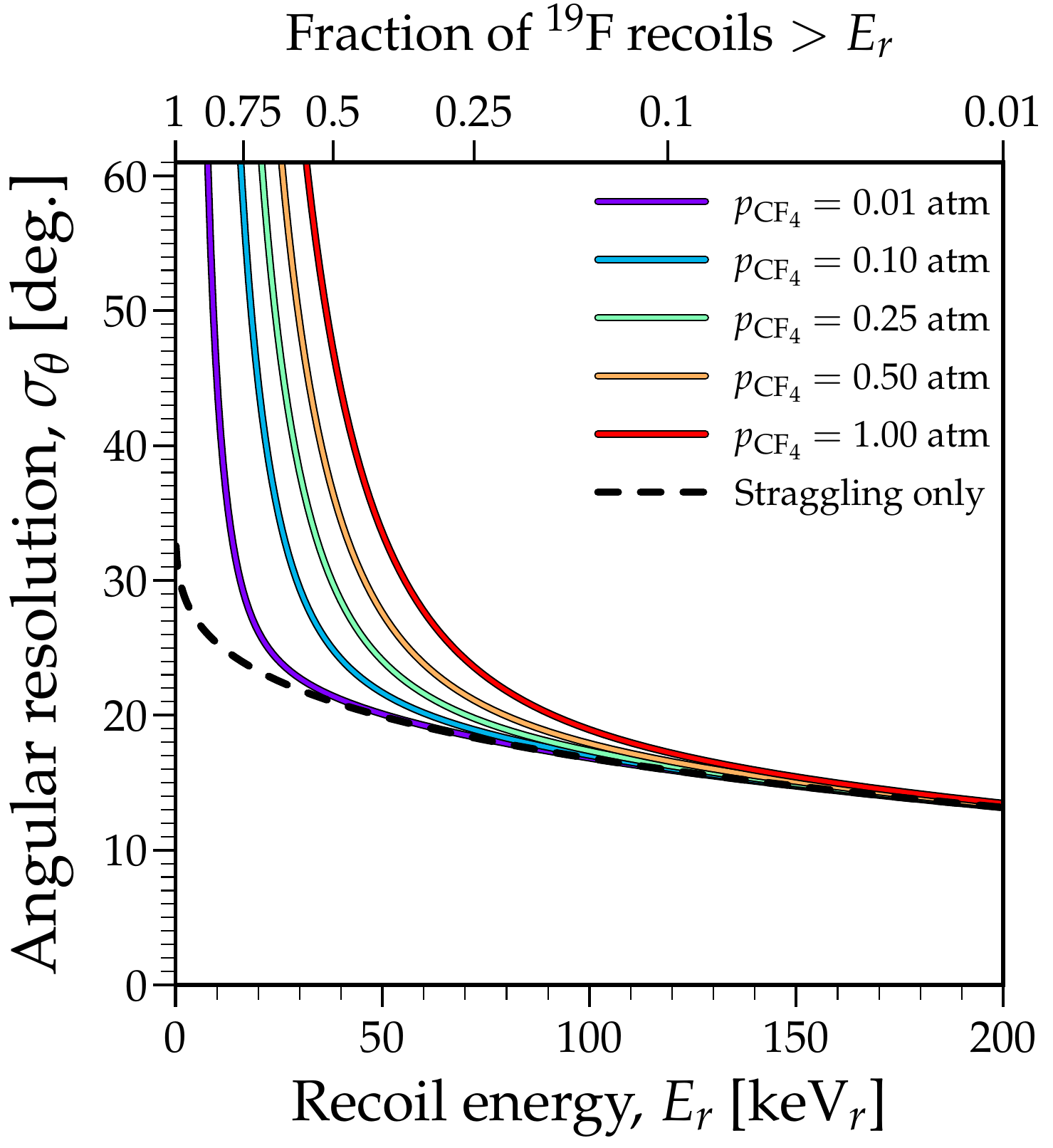}
    \caption{Angular resolution model for our conceptual detector design as a function of nuclear recoil energy, $E_r$, and for five pressures of pure \cffour gas. This model is expressed mathematically in Eq.~(\ref{eq:angularresolution}). The pressure-independent contribution from nuclear straggling alone is shown as a black dashed line. On the upper horizontal axis, we also show the fraction of $^{19}$F recoils lying above the corresponding recoil energy shown on the bottom axis.}
    \label{fig:angular_resolution}
\end{figure}
The uncertainty in the nuclear recoil direction measurement has three main contributions: transverse straggling, diffusion, and detector segmentation. Straggling denotes the scattering of the nuclear recoil, which results in an angular spread of the deposited ionisation. This mechanism dominates at the highest recoil energies, $\mathcal{O}(200)$~keV, where recoils are long compared to the ionisation diffusion scale. At low recoil energies, the recoil length becomes so short that the direction is smeared out by the subsequent diffusion of the ionisation as it drifts in the TPC. This mechanism limits angular resolution at low recoil energies, $\mathcal{O}(10)$~keV. At intermediate recoil energies, the detector segmentation can also contribute if it is not sufficiently small. To reach a conclusion that depends less on the detector specifics, we assume the detector has been designed so that this term is subdominant and negligible. 

Ignoring the detector segmentation, and assuming the total angular resolution is small, the two remaining effects are Gaussian and can be added in quadrature to obtain the total Gaussian angular resolution,
\begin{equation}\label{eq:total_res}
\sigma_\theta(E_r,p) = \sqrt{\sigma_d^2(E_r, p)+\sigma_s^2(E_r)}\,,
\end{equation}
where $\sigma_d$ and $\sigma_s$ are the resolution smearing due to diffusion and straggling. We will model these effects at a reference pressure, $p_0=30$~Torr of CF$_4$ ($\approx 0.04\,{\rm atm}$) so as to compare directly with simulations, and then scale to other gas pressures using known gas detector scaling rules.

There are several details that must be carefully addressed to obtain an accurate model for the two terms in Eq.~(\ref{eq:total_res}): the number of angles that are measured for each recoil, whether a signed 3D vector or an axis is measured, and the fact that our model applies to mismeasurements on a sphere, rather than a flat geometry. We will address the angle count and exact definition first, then build up a Gaussian resolution model, and finally translate this model to a spherical geometry.

First, we assume a detector that independently measures two orthogonal angles for each recoil by combining $N$ measurements of three-dimensional space points, to obtain a 3D recoil direction. In that limit, each of the two measured angles will have an angle-error distribution with a standard deviation given by,
\begin{equation}
\sigma_d^2(E_r,p)= \frac{12 \,z \,\sigma_f^2(p)}{L^2(p,E_r)N(E_r)}\,,
\end{equation}
where $\sigma_f$ is the one-dimensional diffusion seen by each ionised electron during drift in the TPC in the Cartesian coordinate that determines the track angle at hand; $z$ is the distance over which the charge is drifted; and $L$ is the true recoil length (before diffusion). This equation is a back-of-the-envelope prediction but has been validated by comparison with measurements of alpha particles in a TPC with a highly-segmented pixel readout~\cite{Vahsen:2014fba}.

We note that directional detectors often distinguish between 3D axial directionality and 3D vector directionality, where the latter also involves measuring the sign of the recoil vector. Although this sign can be obtained in TPCs from the detailed ionisation distribution along the recoil, it is not straightforward to predict the performance of this sign measurement for all recoil energies and gas pressures. In this analysis, we instead simply assume the signs of all recoil vectors (including background events) point towards a single hemisphere whose pole aligns with the direction to the neutrino source. Then we only need be concerned with the unsigned recoil track axis and model the recoil distribution as a function of $|\cos\theta|$ instead of $\cos\theta$. The error model we are developing, starting from Eq.~(\ref{eq:total_res}), is therefore understood to be predicting the three-dimensional angle error of an (unsigned) axis direction. 

We furthermore assume an ideal detector, which measures each primary electron separately, so that $N=E_rQ(E_r)/W$, where $W$ is the mean energy required to produce one electron–ion pair in the gas at hand. This approximation is the same as was previously used to model detection of electron recoils~\cite{Ghrear:2025iry}. Detecting individual electrons is not realistic for a cost-optimised detector, but should be considered a performance limit for a costly detector. The benefit here is that this choice provides a performance estimate that does not depend on detailed technology choices.

Finally, we assume equal transverse and longitudinal diffusion of ionisation in the TPC, so that the modelled angular resolution becomes isotropic. For \cffour, the longitudinal diffusion in the regime of interest is smaller than the transverse diffusion, but to keep the model simple while remaining conservative, we use the larger transverse diffusion in our estimates below.

We use the software package Stopping and Range of Ions in Matter (SRIM)~\cite{Ziegler:2010bzy} to simulate fluorine recoils in \cffour gas at $p_0=30$~Torr to obtain the linear coefficient $b$ relating the mean true recoil length $L_0$ to the recoil energy $E_r$ via $L_0(E_r) \equiv L(E_r,p_0)=b E_r$. Then finally we extrapolate to other gas pressures using \mbox{$L(E_r,p)=L_0(E_r)p_0/p$} and $\sigma_{f}=\sigma_{f0}\sqrt{p_0/p}$. These two scaling laws are commonly used in the field, and we have confirmed their validity using SRIM and the gas simulation software Magboltz~\cite{Biagi:1999nwa}.

We use the same $p_0=30$~Torr SRIM simulations to obtain the straggling contribution to the angular resolution versus energy, $\sigma_s^2(E_r)$. Note that using two spherical-coordinate angles would involve the Jacobian on a sphere, and make this angular resolution coordinate dependent. We bypass this problem by extracting the 3d angle between the true (unsigned) axis of each simulated recoil direction and the (unsigned) axis of the ionisation recreated by the recoil. The simulated recoil direction is an input to the event generator, while the ionisation track axis is obtained by applying singular value decomposition (SVD) to the ionised electrons obtained from SRIM. We obtain the following fit for the energy-dependence of the mean 3d angle $\alpha$ between the recoil axis and the track axis:
\begin{equation}
    \frac{\langle \alpha \rangle}{1 \,{\rm deg.}} = \frac{1805.23}{\sqrt{26.29 + (E_r/{\rm 1\,keV_r})^{0.291}}} - 307.89 \, .
\end{equation}
The angle $\alpha$---by construction a positive quantity---follows a Rayleigh distribution, and its mean $\langle \alpha \rangle$ is related to the variance of the one-dimensional unsigned axis error distribution, $\sigma^2_s$ in Eq.~(\ref{eq:total_res}) by,
\begin{equation}
\sigma_s = \langle \alpha \rangle \sqrt{\frac{2}{\pi}} \, .
\end{equation}
This relation has been confirmed to hold for our simulation, and establishes a simple methodology for extracting $\sigma_s^2(E_r)$ for recoils pointing in any direction, bypassing the need for a local coordinate system.

Combining all the above, we obtain the following Gaussian model for the angular resolution of fluorine recoils in pure \cffour gas,
\begin{equation}\label{eq:angularresolution}
\sigma_\theta^2(p,E)=
\frac{12\,W\,z\,\sigma_{f0}^2}{b^2\,Q(E_r)E_r^3}
\left(\frac{p}{p_0}\right)+ \sigma_{s}^2(E_r)\,,
\end{equation}
where $p_0= 30~\mathrm{Torr}$ is the reference pressure of \cffour, $b = 0.00452~\mathrm{cm}\,\mathrm{keV}_r^{-1}$ and $W=0.034$~keV. We use $z = 25\,{\rm cm}$ to represent the average drift length and $\sigma_{f0} = 340\,\upmu \mathrm m/\sqrt{\mathrm{cm}}$ for the diffusion, obtained by scanning the electric field strength for the minimum diffusion in 30~Torr of \cffour, using Magboltz.

We plot this energy- and pressure-dependent angular resolution in Fig.~\ref{fig:angular_resolution}. The angular performance degrades sharply at low recoil energies. For these low-energy tracks, the direction assignment would be essentially random, and the angular distribution would asymptote to one that was indistinguishable from an isotropic distribution of axes. For pressures around half an atmosphere, the sharp degradation in the performance takes over for recoil energies below $\sim$40 keV. Reading the upper axis in this figure, we see that this would imply that around 60\% of recoils would have reconstructed track directions correlating at some level with the initial (true) recoil direction, although only those above $\sim$50~keV would correlate well enough to measure the angular distribution of \cevns recoils, i.e.~$\sigma_\theta\lesssim 26$--$30^\circ$ (twice the standard deviation of the underlying angular distribution). We can use an argument of this kind to estimate the maximum tolerable pressure at which we still have an acceptable level of directionality across the majority of the \cevns recoil spectrum.

\begin{figure}
    \centering
  \includegraphics[width=0.99\linewidth]{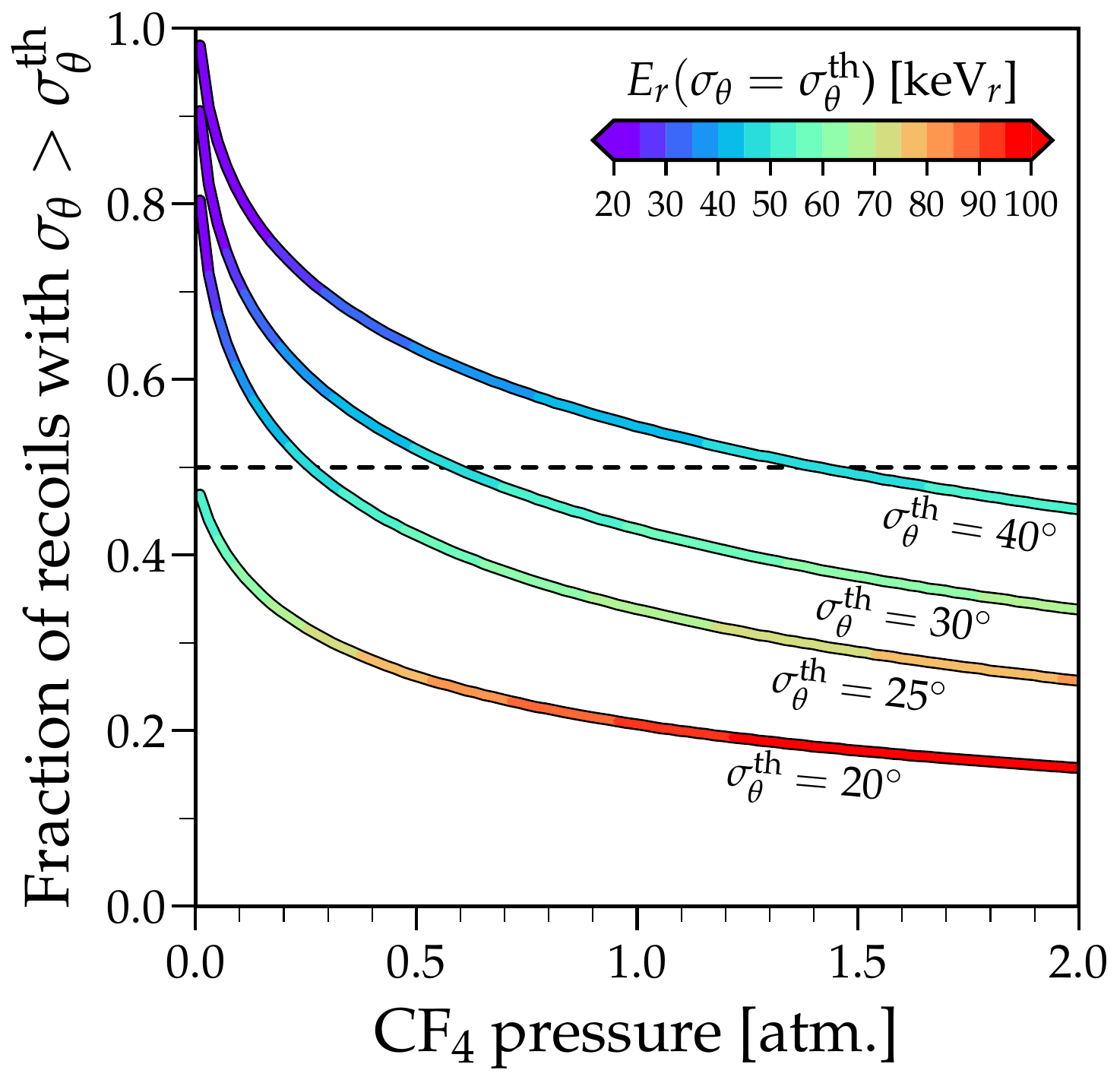}
    \caption{Fraction of events with angular resolution better than some ``directionality threshold'', $\sigma_\theta<\sigma_\theta^{\rm th}$. We show this fraction as a function of the \cffour pressure since the angular resolution degrades for higher gas densities. We show four example choices for $\sigma_\theta^{\rm th}$. If we wish the majority of the sample of events to have angular resolution better than $\sigma_\theta<26$--$30^\circ$ (twice the standard deviation of the true underlying recoil angle distribution), then a gas pressure lower than 0.4--0.5 atmospheres is required.}
    \label{fig:DirectionalityThreshold}
\end{figure}
To demonstrate this, we present in Fig.~\ref{fig:DirectionalityThreshold} the fraction of all detected events lying above some ``directionality threshold'' $\sigma_\theta^{\rm th}$ as a function of the pressure. If we wish the majority of the detected \cevns events to have good-enough angular resolution, then this implies a choice of \cffour pressure around 0.3--0.5 atm. A lower pressure than this would improve the directionality for more of the detected events, but this would also reduce the total event rate statistics, which scales linearly with $p_{\rm CF_4}$. On the other hand, we could obtain a higher event rate using a higher pressure, but then only a minority of the detected tracks would have reliable directionality, defeating the purpose of this measurement. So we will settle on a pressure around this value for this study, keeping in mind that this is a flexible parameter that the rest of the experimental design does not depend upon. In fact, it would be possible to run multiple iterations of the same experiment in a high-statistics/low-directionality mode with pressures close to atmospheric to obtain a high-significance measurement of the rate, as well as a low-statistics/high-directionality mode, which sacrifices the event rate in favour of more precise recoil direction measurements. Our fiducial \cffour pressure of 0.4 atm for this study is chosen simply to estimate the performance that lies at a balanced midpoint between these two possible scenarios.

\subsection{Recoil distributions}\label{sec:recoildistributions}
\begin{figure*}
    \centering
    \includegraphics[trim={0, 2cm, 0, 0},clip,width=0.99\linewidth]{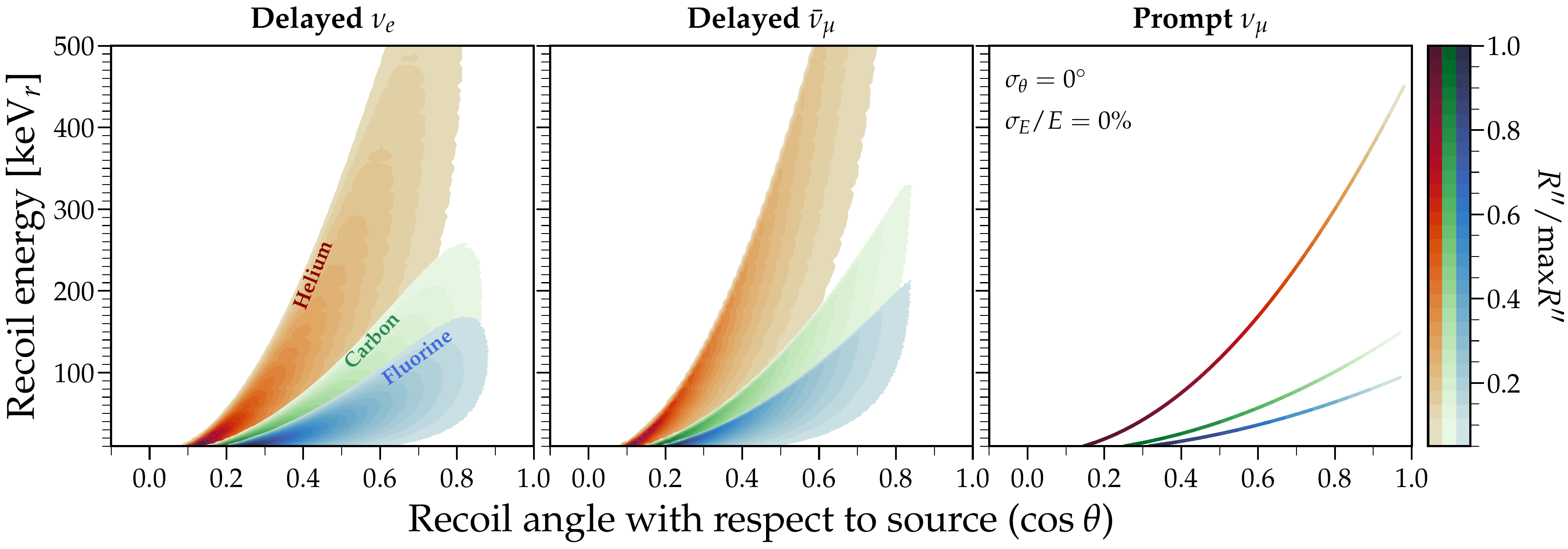}
    \includegraphics[width=0.99\linewidth]{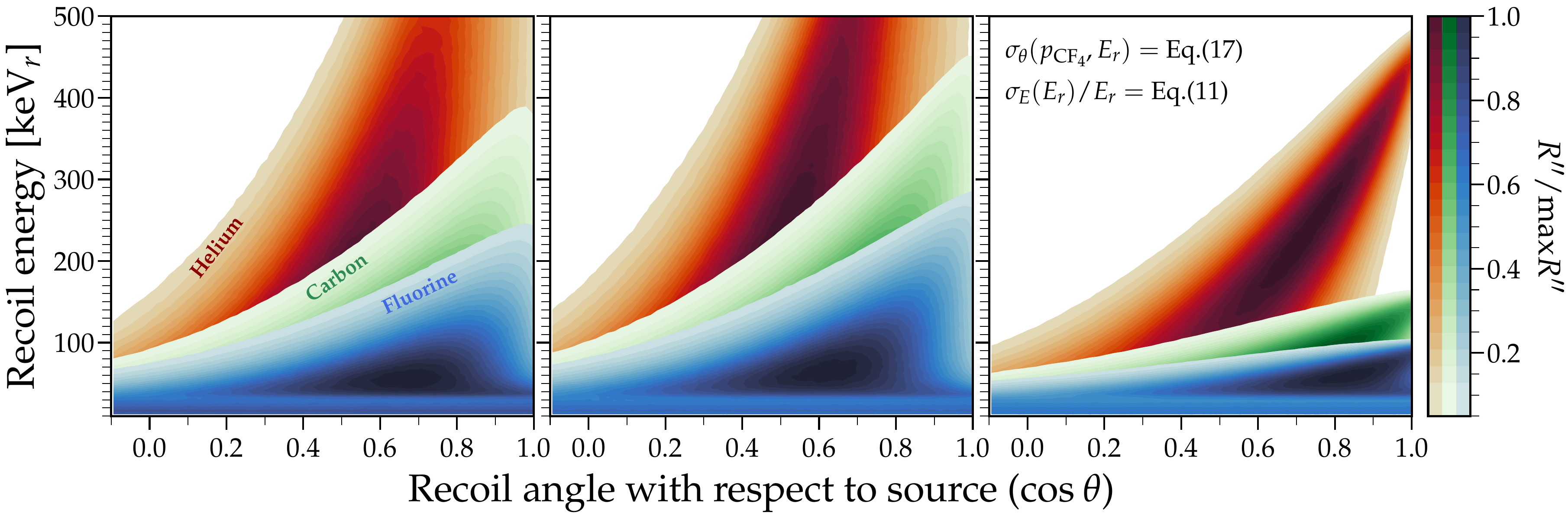}
    \caption{Recoil energy and angle distributions, plotted as $R^{\prime \prime}/{\rm max}(R^{\prime \prime})$ where $R^{\prime \prime} = {\rm d}^2R/{\rm d}E_r{\rm d}\Omega_r$, as in Eq.~(\ref{eq:d2RdEdOmega}). From left to right, we show the distributions for the three neutrino fluxes: the continuous delayed fluxes of $\nu_e$ and $\bar{\nu}_\mu$, and the mono-energetic prompt flux of $\nu_\mu$. The red, green, and blue contours correspond to the three nuclear species in our chosen gas mixture. The top set of panels shows the underlying (theoretical) recoil distributions, while the bottom panels show the distributions after incorporating the finite energy and angle reconstruction performance, parameterised using Eq.~(\ref{eq:energyresolution}) and Eq.~(\ref{eq:angularresolution}). The bottom set of panels (after summing over the three nuclei and three fluxes) constitutes our signal model for our sensitivity analysis described in Sec.~\ref {sec:SM}.}
    \label{fig:RecoilDistributions}
\end{figure*}
Before moving on to our sensitivity analysis, we describe how we implement the finite angular performance in our Monte Carlo simulation to obtain our final measured recoil distributions. The parameterisation of $\sigma_\theta$ we have introduced so far has assumed that each of the two mismeasurements it models are small and add in quadrature. However, an essential part of our analysis is the regime at low recoil energies and high gas pressure, where mismeasurements become large and angular sensitivity vanishes. The strict Gaussian model and addition in quadrature are not valid in this limit because the Gaussian distribution ignores the fact that the angles exist on a sphere. However, the Gaussian variance can be translated into the width parameter $\kappa$ of a von Mises-Fisher (vMF) distribution, which can be thought of as a (symmetric) Gaussian distribution on a sphere. We use a Monte Carlo simulation to derive the exact relationship between the two, which is approximately,
\begin{equation}
\kappa \approx \frac{1}{(\sigma_\theta/1\,{\rm rad})^2}\,.\label{eq:sigma_to_kappa}
\end{equation}
We include the above effects of a finite track direction reconstruction by first sampling initial `true' track directions, $\hat{\mathbf{q}}$ from $\textrm{d}^2R/\textrm{d}E_r\textrm{d}\Omega_r$ and then re-sampling `measured' directions $\hat{\mathbf{q}}^\prime$ from the vMF distribution,
\begin{equation}
K_{\theta}(\hat{\mathbf{q}}, \hat{\mathbf{q}}^\prime,E_r)=\frac{\kappa(E_r)}{4 \pi \sinh \kappa(E_r)} \exp \left(\kappa(E_r) \mathbf{q} \cdot \mathbf{q}^\prime \right) \,.
\end{equation}
This procedure allows us to generate Monte Carlo events with proper mismeasurements, even in the regime where the Gaussian model tails would start to wrap around the sphere.

We visualise the resulting recoil distributions as a function of our two main observables in Fig.~\ref{fig:RecoilDistributions}. The underlying \textit{theoretical} event distribution as a function of $E_r$ and $\cos\theta$ is expressed by $\drm^2 R/\drm E_r \drm \Omega_r$, as introduced Sec.~\ref{sec:cevns}, is shown on the top three panels. The `measured' distribution shown in the bottom three panels is computed using the Monte Carlo procedure described above, which applies the effects of the detector's finite ability to reconstruct the recoil energy and angle. The three columns correspond to the different neutrino fluxes ($\nu_e$, $\bar{\nu}_{\mu}$ and $\nu_\mu$) and we use three different colours in each case to show how the distributions depend on the nucleus that is recoiling (although we emphasise our event rate will always be dominated by fluorine recoils due to the $\sim N^2$ dependence of the \cevns cross section).

The two delayed fluxes are spread over a range of neutrino energies: $E_\nu \in [0,m_\mu/2]$~MeV, which means they lead to a spread of recoils occupying the kinematically allowed space. This varies for each nucleus because of the dependence on $m_N$ in the scattering kinematics, cf.~Eq.~(\ref{eq:costh}). The prompt $\nu_\mu$ flux, however, is mono-energetic at $E_\nu \approx 30$~MeV, which means events appear along a one-dimensional subspace of the two-dimensional ($E_r,\,\cos\theta$) plane, defined by Eq.~(\ref{eq:costh}). Once the energy and angular resolutions are taken into account the distributions become much less distinct, as is expected, however the kinematic space occupied by each nucleus' distribution (and to some extent each neutrino flux) are still partially separated, which will enable the directional information to provide additional statistical power when it comes to making measurements of the target-dependent neutrino cross sections or the three fluxes, when compared to an equal-standing non-directional measurement. We will quantify this statement in Sec.~\ref{sec:SM}.

\subsection{Event-by-event neutrino energy measurements}\label{sec:energyreconstruction}
\begin{figure*}
    \centering
    \includegraphics[width=0.96\linewidth]{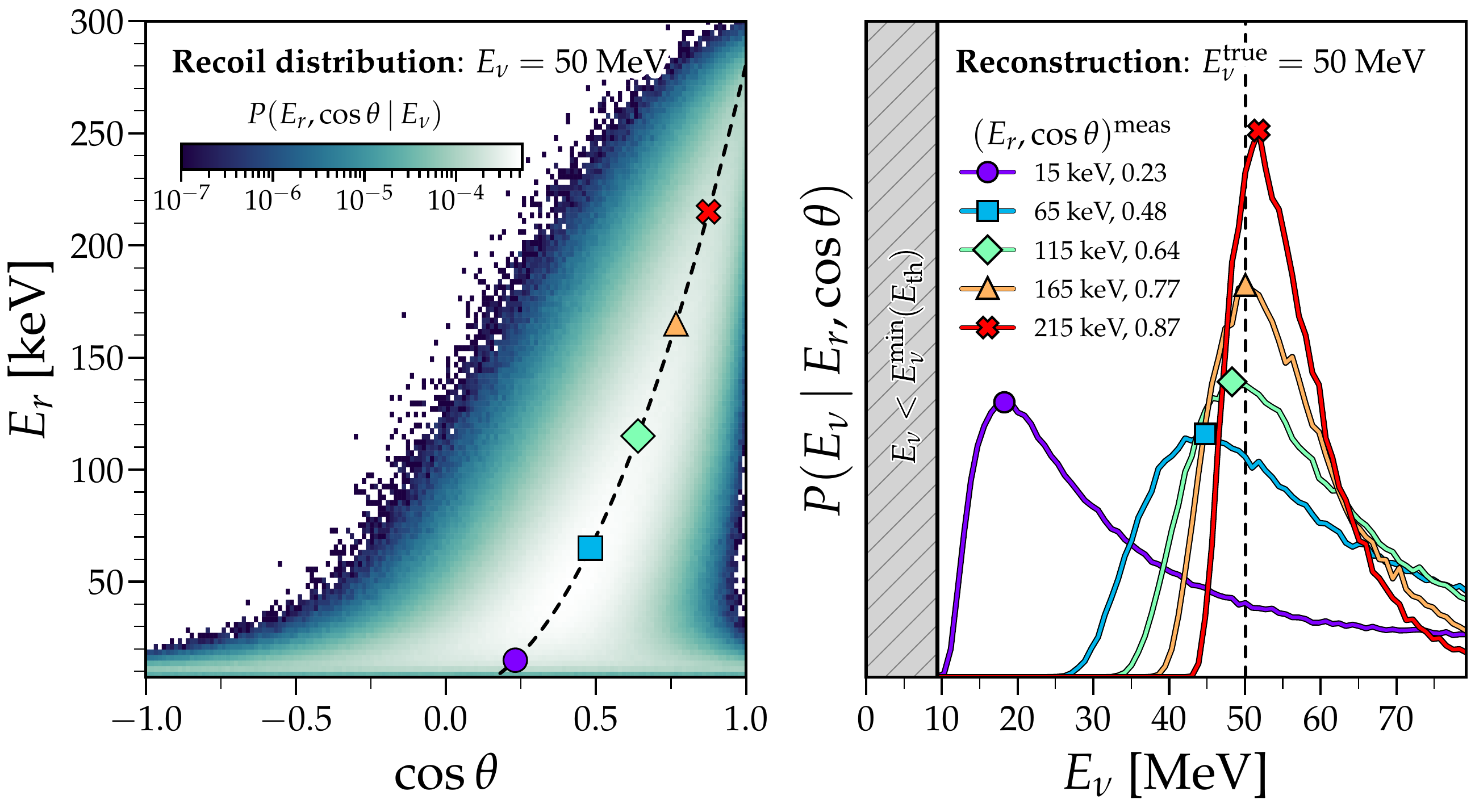}
    \caption{Visualisation of the event-by-event reconstruction of the neutrino energy. {\bf Left}: the distribution of \textit{measured} recoil energies and $\cos\theta$ angles given a fixed initial neutrino energy $E_\nu = 50$~MeV. We express this distribution as a joint probability distribution, $P(E_r,\cos\theta\,|\,E_\nu)$. Because we assume a fixed neutrino energy, the `true' distribution sits along a one-dimensional subspace indicated by the dashed line. We overlay five arbitrary benchmark recoil energy and angle pairs with coloured symbols. {\bf Right}: The distribution of \textit{inferred} $E_\nu$ given some pair of measured $(E_r,\cos\theta)$, i.e.~the probability distribution $P(E_\nu \, | E_r,\cos\theta)$, which we calculate by computing the full distribution $P(E_r,\cos\theta, E_\nu)$ over a wide range of neutrino energies and assuming the SM \cevns cross section. We choose $(E_r,\cos\theta)$ pairs marked with coloured symbols in the left-hand plot, which lie along the $E_\nu = 50$~MeV line. The most-probable value of $E_\nu$ converges on $50$~MeV only for large $E_r$, where the energy/angular resolutions are good. We shade in gray the region where $E_\nu<E_\nu^{\rm min}(E_{\rm th})$ for $E_{\rm th} = 10$~keV.}
    \label{fig:NeutrinoEnergyReconstruction}
\end{figure*}
\begin{figure*}
    \centering
    \includegraphics[width=0.45\linewidth]{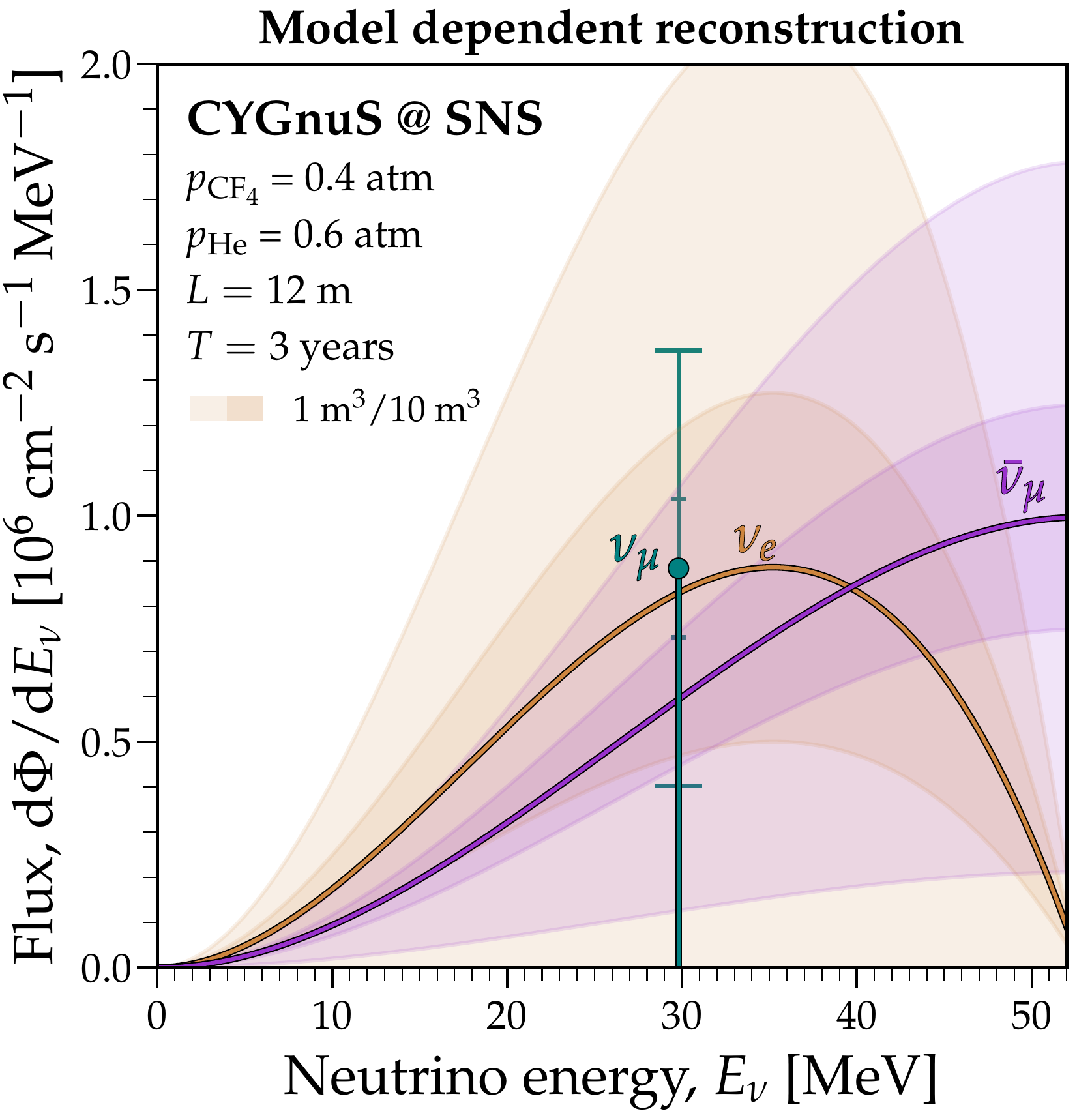}\hspace{3em}
    \includegraphics[width=0.45\linewidth]{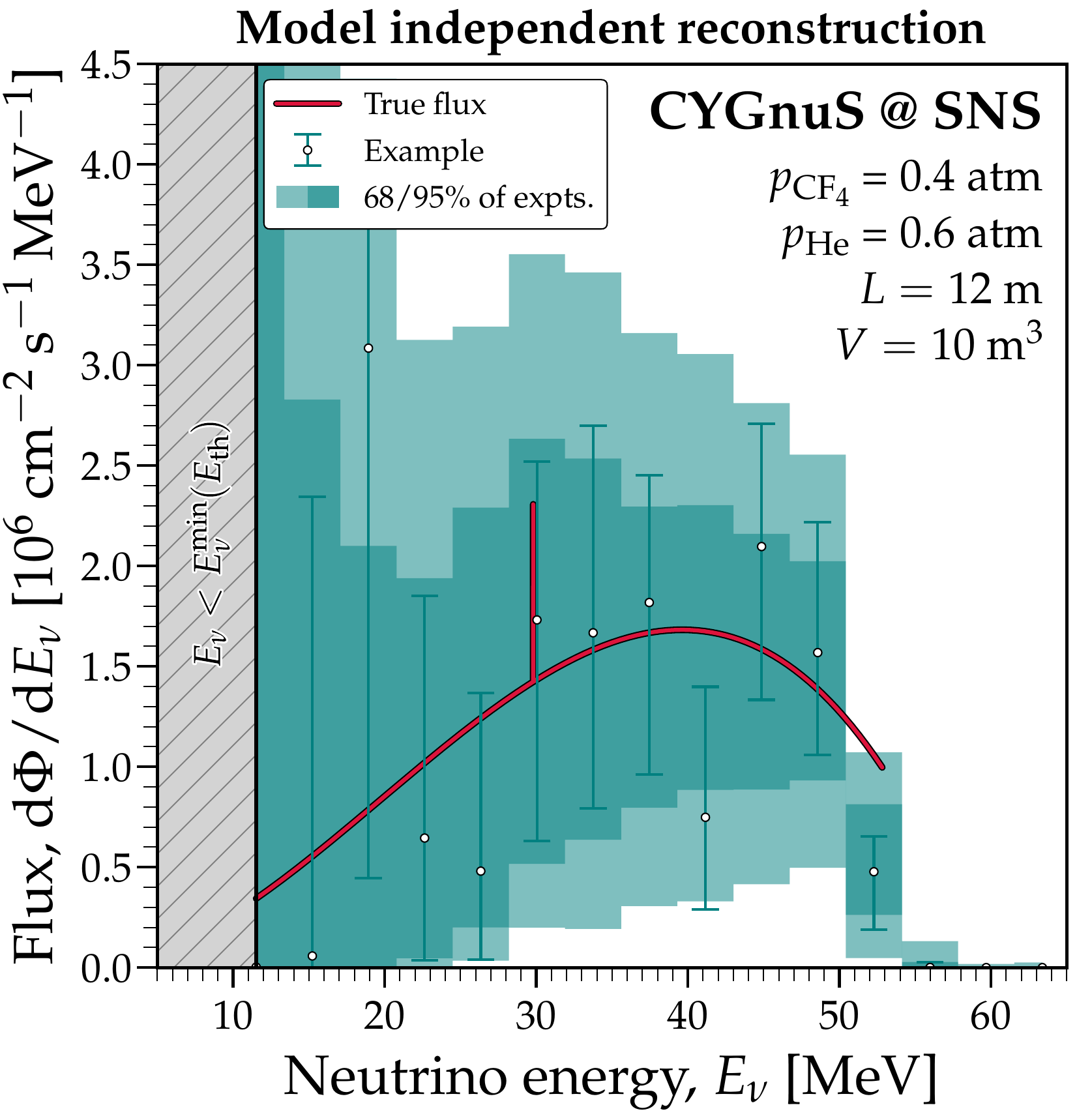}
    \caption{Two complementary methods for reconstructing the neutrino energy spectrum and flux. {\bf Left}: We show the expected (median) $1\sigma$ uncertainty bounds around each neutrino flux component obtained from our profile-likelihood ratio analysis (Sec.~\ref{sec:likelihood}). The case for 1 and 10 m$^3$ experiments are shown by the lighter and darker transparency bands respectively. For the $\nu_\mu$ line, the wider error bar corresponds to the 1 m$^3$ case and the shorter one corresponds to the 10 m$^3$. The true fluxes are shown by solid lines. {\bf Right}: expected ``model-independent'' reconstruction of the flux from the event-by-event neutrino energy reconstruction method described in Sec.~\ref{sec:energyreconstruction}. The error bars show the result obtained by a randomly generated set of data in 10 m$^3$, while the shaded bands give the 68 and 95\% containment around all possible hypothetical experiments. The true underlying flux (the sum of the three fluxes in this case, since there is no flavour information retained by the recoils themselves) is shown by a red line. It is not possible to reconstruct the flux for energies $E_\nu<E_\nu^{\rm min}(E_{\rm th})$ which is shown by the gray region.}
    \label{fig:NeutrinoSpectrumReconstruction}
\end{figure*}

One of the most attractive prospects for a direction-sensitive neutrino recoil detector is the ability to reconstruct the initial neutrino energy spectrum, without relying on a model for it. The reason this is possible is because the neutrino source is point-like and so the kinematic expression Eq.~(\ref{eq:costh}) implies a one-to-one relationship between a pair of $(E_r,\,\cos\theta)$  and the neutrino energy $E_\nu$ responsible for them. This makes possible, in principle at least, event-by-event neutrino energy reconstruction.

The situation is made slightly more difficult by the fact that we cannot measure $E_r$ and $\cos\theta$ to arbitrary precision. This means the reconstruction will not be perfect, but it will still be possible to some extent, which we will now show. In Fig.~\ref{fig:NeutrinoEnergyReconstruction} we show an example that illustrates how the neutrino energy reconstruction may work in practice. The left-hand plot shows a probability distribution of recoils (similar to the bottom panels of Fig.~\ref{fig:RecoilDistributions}) that we express as $P(E_r,\cos\theta \, |\,E_\nu)$, i.e.~it is the distribution of measured recoil energies and angles given a fixed value for $E_\nu$, assumed here to be 50~MeV. On the right, we show the inverted conditional probability $P(E_\nu\,|\,E_r,\cos\theta)$, i.e.~how probable is the neutrino energy $E_\nu$ for some measured pair of $(E_r,\cos\theta)$ given the SM neutrino cross section.

We have chosen the benchmark values of $(E_r,\cos\theta)$ to lie along the correct kinematic line on the left-hand plot for $E_\nu = 50$~MeV. If the measurement of $E_r$ and $\cos\theta$ were perfect (i.e. $\sigma_\theta = \sigma_E = 0$), then the probability distributions on the right-hand side would all be delta functions centred at $E_\nu = 50$~MeV. However, we see that the distributions are broadened and also skewed due to mismeasurement of the recoil energy and angle. The non-Gaussianity in this probability distribution originates because of the combined energy-dependence of both the angular and energy resolution functions, as well as the kinematic constraint---namely, certain values of $E_\nu$ will not be returned by Eq.~(\ref{eq:costh}) for a given $(E_r,\cos\theta)$ if they are not permitted by energy and momentum conservation.

The most striking observation in the right-hand panel of Fig~\ref{fig:NeutrinoSpectrumReconstruction} is the fact that the most probable neutrino energy does not align with $E_\nu = 50$~MeV for low values of the measured recoil energy. To understand this, consider the fact that to compute $P(E_\nu\,|\,E_r,\cos\theta)$ we assumed the SM \cevns cross section, which decreases linearly with increasing recoil energy for a fixed neutrino energy, i.e.~$\textrm{d}\sigma/\textrm{d}E_r \propto (1-m_N E_r/2E_\nu^2)$. This means that for a given measured $E_r$, if one has no prior knowledge of $E_\nu$, it is more likely that the neutrino energy is at lower energies than higher energies due to this dependence in the cross section. The independent measurement of $\cos\theta$ is supposed to resolve this degeneracy, but at very low recoil energies, the angular resolution is very large, and so we have almost no ability to measure $\cos\theta$. This can be seen in the left-hand side of Fig.~\ref{fig:NeutrinoEnergyReconstruction}, where the recoil distribution is spread across the full domain from $\cos\theta = -1$ to $1$ at low energies. 

Each measured pair of recoil energy and recoil angle allows one to unfold the two-dimensional distribution to obtain a probability distribution for the corresponding neutrino energy. We can then combine these together for a larger sample of recoil events to reconstruct the original distribution of neutrino energies. We will use this approach in the next section to illustrate how a directional detector could make a `model-independent' reconstruction of the flux.

\section{Standard Model measurements}\label{sec:SM}
We will now describe our statistical formalism for deriving sensitivity projections for several potential measurements of interest, assuming, for the time being, no additional neutrino-nucleus interactions beyond the Standard Model.

\subsection{Background model}
Although our goal in this study is not to perform a detailed background assessment, the level of backgrounds, especially nuclear recoils induced by neutrons, will be an important factor influencing our sensitivity. So we aim here to adopt a conservative model by assuming our background rate of events is comparable to the rate of neutrino events within each pulse (i.e. a signal purity of $R_{\rm \nu}/(R_\nu + R_{\rm bg}) = 0.5$). 

The differential recoil spectrum for our background model is set to the following simplified form,
\begin{equation}
    \frac{\textrm{d}^2 R_{\rm bg}}{\textrm{d} E_r \textrm{d}\Omega_r} = \frac{R_{\rm bg}}{4\pi E_{\rm bg}(e^{-E_{\rm th}/E_{\rm bg}} - e^{-E_{\rm max}/E_{\rm bg}})} e^{-E_r/E_{\rm bg}} \, .
\end{equation}
This model is isotropic in angle and exponentially falling in energy, with $E_{\rm bg} = 67$~keV, which is essentially a worst-case scenario because the spectrum then overlaps with most of the expected \cevns recoil spectrum and returns the same mean recoil energy. Although we do anticipate the background to be exponentially falling, the spectral shape would require measurements to determine, which could be performed if the time profile of neutron events was distinct from the neutrinos. The background's angular distribution is also unlikely to be perfectly isotropic, as there will be a source of neutrons originating from the same direction as the neutrinos, in addition to cosmics, which would emerge from overhead. However, the large amount of shielding anticipated for this experiment would cause much of the neutron background to isotropise. Moreover, the assumption of isotropy can be considered a conservative choice, as any variation in the angular distribution that was distinct from the neutrino distribution would aid in distinguishing them. As a final step, we will pessimistically float the background model as a free parameter in our likelihood function described below, as this will allow us to demonstrate that the addition of directional information provides another major advantage in the form of excellent background rejection.

\subsection{Likelihood framework}\label{sec:likelihood}
Our likelihood is based on two-dimensional binned pseudo-data, with bins running over the space $(E_r,|\cos{\theta}|)$. Given an observed number of events in each bin $N^{kl}_{\rm obs}$ and an expected number $N^{kl}_{\rm exp}(\boldsymbol{\theta})$, the likelihood is a function of some set of model parameters, $\boldsymbol{\theta}$. It is constructed from the product of the Poisson probabilities ($\mathscr{P}$) for the expected number of events in each energy and angle bin,
\begin{equation}\label{eq:likelihood}
 \mathscr{L}(\boldsymbol{\theta})= \prod_{k,l} \mathscr{P} \left(N_\textrm{obs}^{kl} \bigg| N^{kl}_{\rm exp}(\boldsymbol{\theta})\right) \, .
\end{equation}
The expected number of events in bin $kl$ is,
\begin{align}
    N_{{\rm exp}}^{k l}(\boldsymbol{\theta})=T & \int_{|\cos \theta^{k}|}^{|\cos \theta|^{k+1}} \int_{E_{r}^{l}}^{E_{r}^{l+1}}\left[\frac{\mathrm{d}^{2} R_{\rm tot}(\boldsymbol{\theta})}{\mathrm{d} E_{r} \mathrm{~d} |\cos \theta|}\right] \nonumber \\
    & \mathrm{~d} |\cos \theta| \mathrm{~d} E_{r} \, ,
\end{align}
where $T$ is the total running time of the experiment, and the total rate is found by summing over each neutrino flux and each target,
\begin{align}
    \frac{\mathrm{d}^{2} R_{\rm tot}(\boldsymbol{\theta})}{\mathrm{d} E_{r} \mathrm{~d} |\cos \theta|} =& f_{\rm bg} M\frac{\drm^2 R_{\rm bg}}{\drm E_r\drm |\cos\theta|} \nonumber \\
    &+ \sum_{i,j}f_{\nu_i} f_{T_j} M_{T_j}\frac{\mathrm{d}^{2} R_{\nu_i,T_j}}{\mathrm{d} E_{r} \mathrm{d}|\cos\theta|} \, .
\end{align}
where $M_{T_j}$ is the mass of the detector composed of target species $j$ and $M = \sum_j M_{T_j}$. For the purposes of introducing our framework, we choose the free parameters of this model to be a series of normalisation constants applying to each neutrino flux and the cross section for each nucleus, i.e.~$\boldsymbol{\theta} = \{\boldsymbol{f}_\nu,\boldsymbol{f}_T,f_{\rm bg}\}$ where $\boldsymbol{f}_{\nu} = \{ f_{\nu_e},f_{\bar{\nu}_\mu},f_{\nu_\mu}\}$ and $\boldsymbol{f}_{T} = \{f_{\rm He},f_{\rm C},f_{\rm F} \}$. To generate our pseudo-data, we assume the ``true'' set of these values is $\forall \boldsymbol{\theta} = 1$. We then construct a statistical test to determine the precision with which these parameters can be measured. We introduce this in general terms here, however in most of our examples, we will not float all of the parameters in $\boldsymbol{\theta}$ at the same time, but rather fix some and let others vary depending on which result we wish to show. We will also introduce new free parameters in Sec.~\ref{sec:BSM} when we introduce BSM interactions and sterile neutrinos.

Using the likelihood written above, we can construct a hypothesis test statistic (TS) out of a likelihood ratio, which compares the maximised likelihood values between two points in the parameter space. The interesting case for us will be to test for the presence of the neutrino signal and exclude the null hypothesis that one or more of the normalisation constants describing the signal are equal to zero. With that in mind, we write down the standard profile likelihood ratio test statistic used in the context of a discovery, i.e.~quantifying the evidence in favour of a positive signal against the null hypothesis of there being no signal~\cite{Cowan:2010js}. We will compute values of TS as a function of both one and two parameters, so to express this we imagine separating out some arbitrary number of parameters from the full set, i.e.~$\boldsymbol{\theta} =\{\boldsymbol{\theta}_0,\boldsymbol{\theta}_1 \}$ where $\boldsymbol{\theta}_0$ are the parameter values of interest, and $\boldsymbol{\theta}_1$ are the rest of the parameters that will be profiled over. The TS is then,
\begin{equation}
    {\rm TS}(\boldsymbol{\theta}_0) = 2\Delta \ln \mathcal{L} = 2\left( \ln \mathcal{L}(\boldsymbol{\theta}_0,\hat{{\boldsymbol{\theta}}}_1) - \ln \mathcal{L}(\mathbf{0},\hat{\hat{\boldsymbol{\theta}}}_1) \right)\nonumber \, ,
\end{equation}
where $\hat{\hat{\boldsymbol{\theta}}}_1$ are the maximum likelihood estimators (MLEs) when all parameters in $\boldsymbol{\theta}_0$ are fixed to zero in the likelihood, whereas $\hat{{\boldsymbol{\theta}}}_1$ are MLEs when $\boldsymbol{\theta}_0$ are fixed to the particular chosen values at which TS is being evaluated. We will assume that Wilks' theorem holds, which states that the distribution of TS asymptotically follows a $\chi^2_{n}$ distribution where $n$ is the number of parameters in $\boldsymbol{\theta}_0$, and so the significance of a particular value of TS can be computed from the $\chi_n^2$ inverse cumulative distribution function.


In practice, the distribution of TS should be calculated using a Monte Carlo procedure to validate (or otherwise) the use of Wilks' theorem. For instance, it is likely to be slightly inaccurate for our 1 m$^3$ case due to the small event numbers. However, since we simply wish to estimate sensitivities for the purposes of optimising the parameters of our experimental setup, conducting the full procedure for finding the correct distribution of TS is an unnecessary complication. So to simplify our approach, we will adopt the commonly-used Asimov dataset technique, which allows us to estimate the median (expected) sensitivity at minimal computational cost~\cite{Cowan:2010js}. This is a hypothetical dataset in which the observation exactly matches the expectation for a given model, i.e.~$N^{kl}_{\rm obs} = N^{kl}_{\rm exp}$ for all $k$ and $l$. It can be shown that the test statistic computed assuming this dataset asymptotes towards the median of the model's TS distribution as the number of observations increases~\cite{Cowan:2010js}.

\subsection{Neutrino fluxes and spectra}\label{sec:measure_flux}
\begin{figure}
    \centering
    \includegraphics[width=0.99\linewidth]{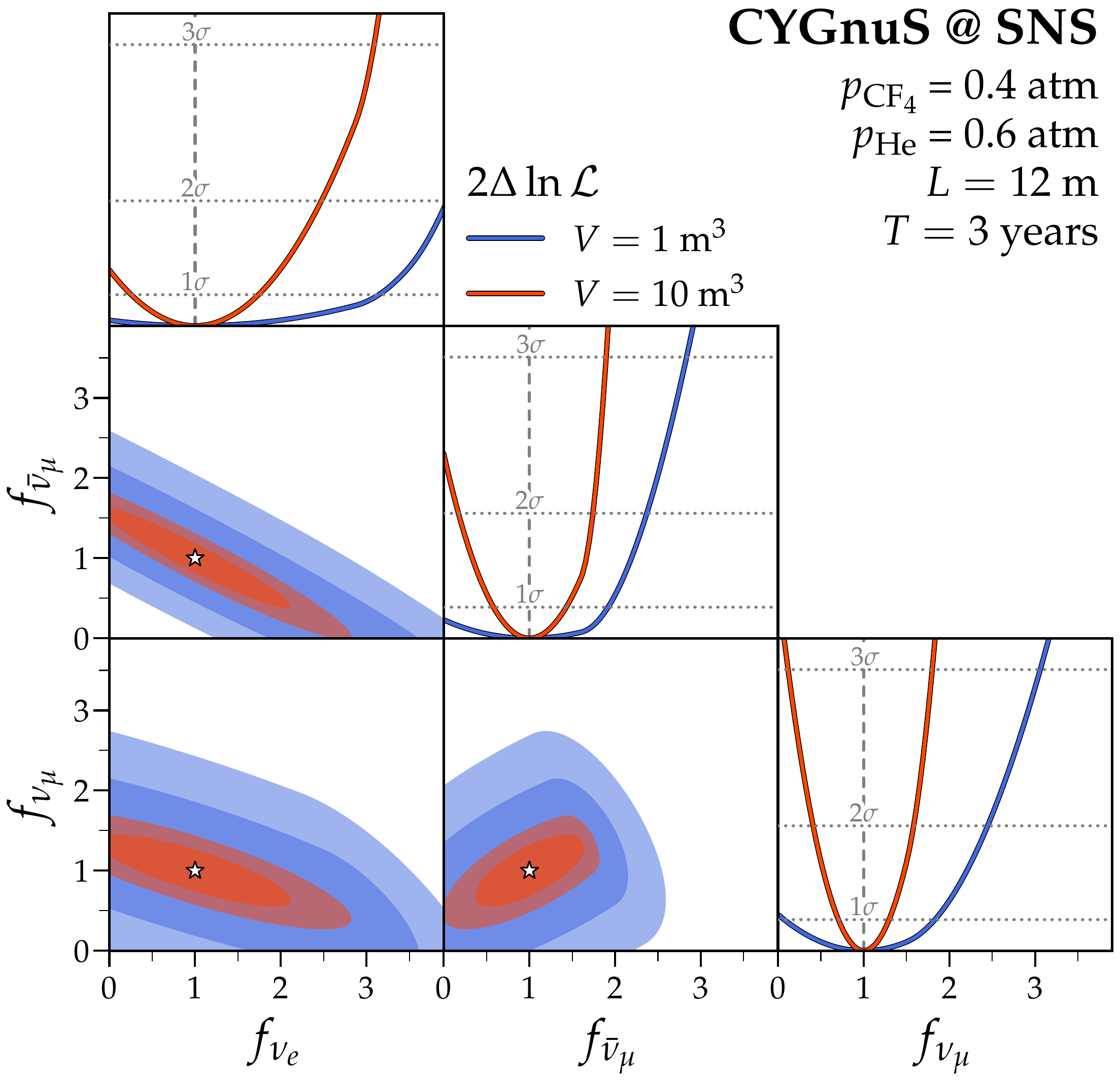}
    \caption{One and two-dimensional Asimov profile likelihood ratios for the reconstruction of the three normalised flux parameters $f_{\nu_i} = \Phi_{\nu_i}/\Phi_{\nu_i}^{\rm true}$. Blue is used for the 1 cubic metre detector, while orange is used for the 10 cubic metre detector. In the two-parameter panels, the two contours are for 1 and 2$\sigma$ significance, while we use horizontal lines for the significance values in the one-parameter panels.}
    \label{fig:NuFluxReconstruction}
\end{figure}

\begin{figure}[t]
    \centering
    \includegraphics[width=0.96\linewidth]{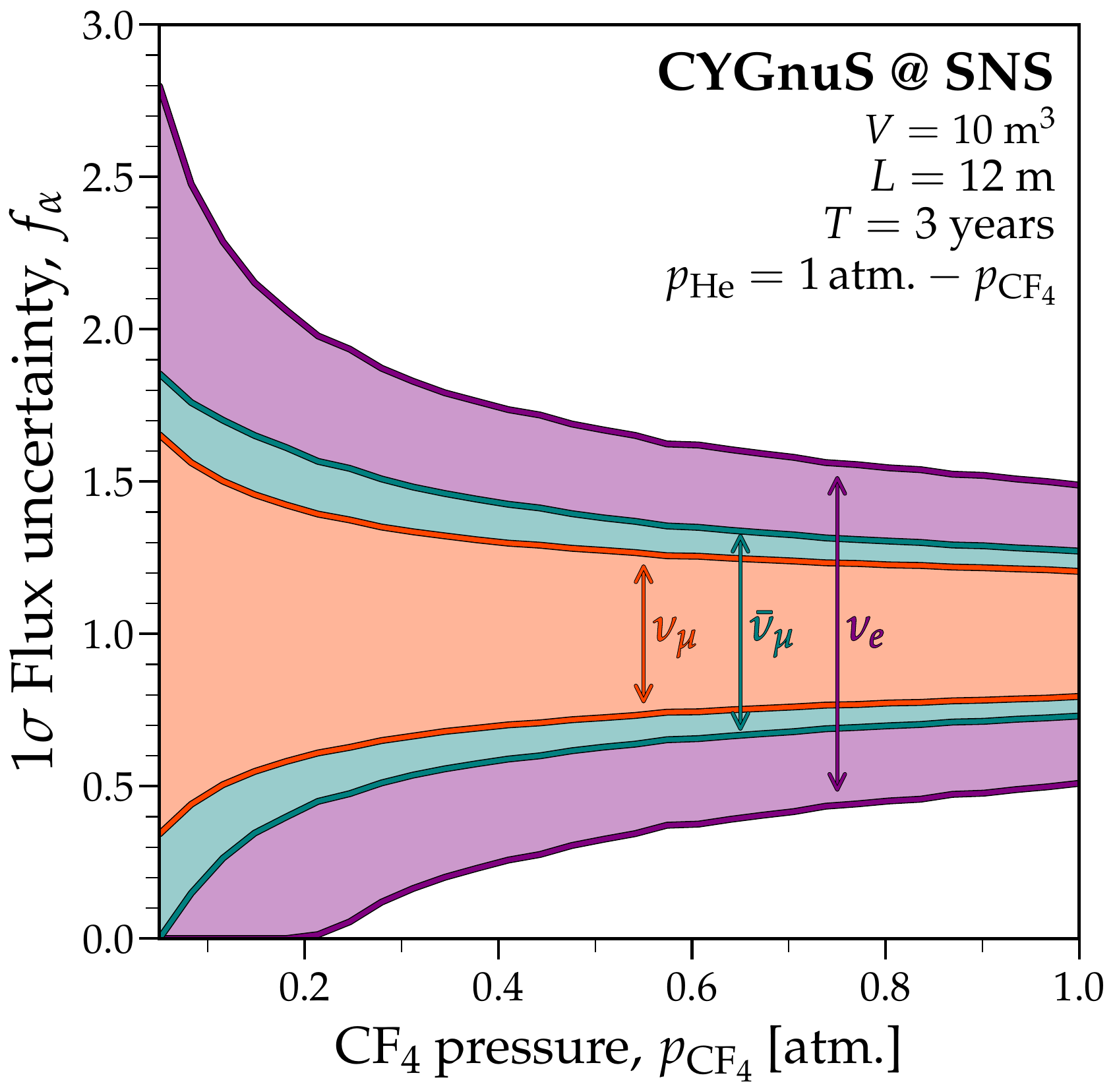}
    \caption{We show the median $1\sigma$ measurement uncertainty from our profile likelihood ratio test for each flux normalisation as a function of the \cffour pressure. Recall that the \textit{total} pressure is always kept to 1 atmosphere, and so the pressure in helium is $p_{\rm He} = 1\,{\rm atm} - p_{\rm CF_4}$.}
    \label{fig:FluxMeasurement_vs_pressure}
\end{figure}
\begin{figure}
    \centering
    \includegraphics[width=0.99\linewidth]{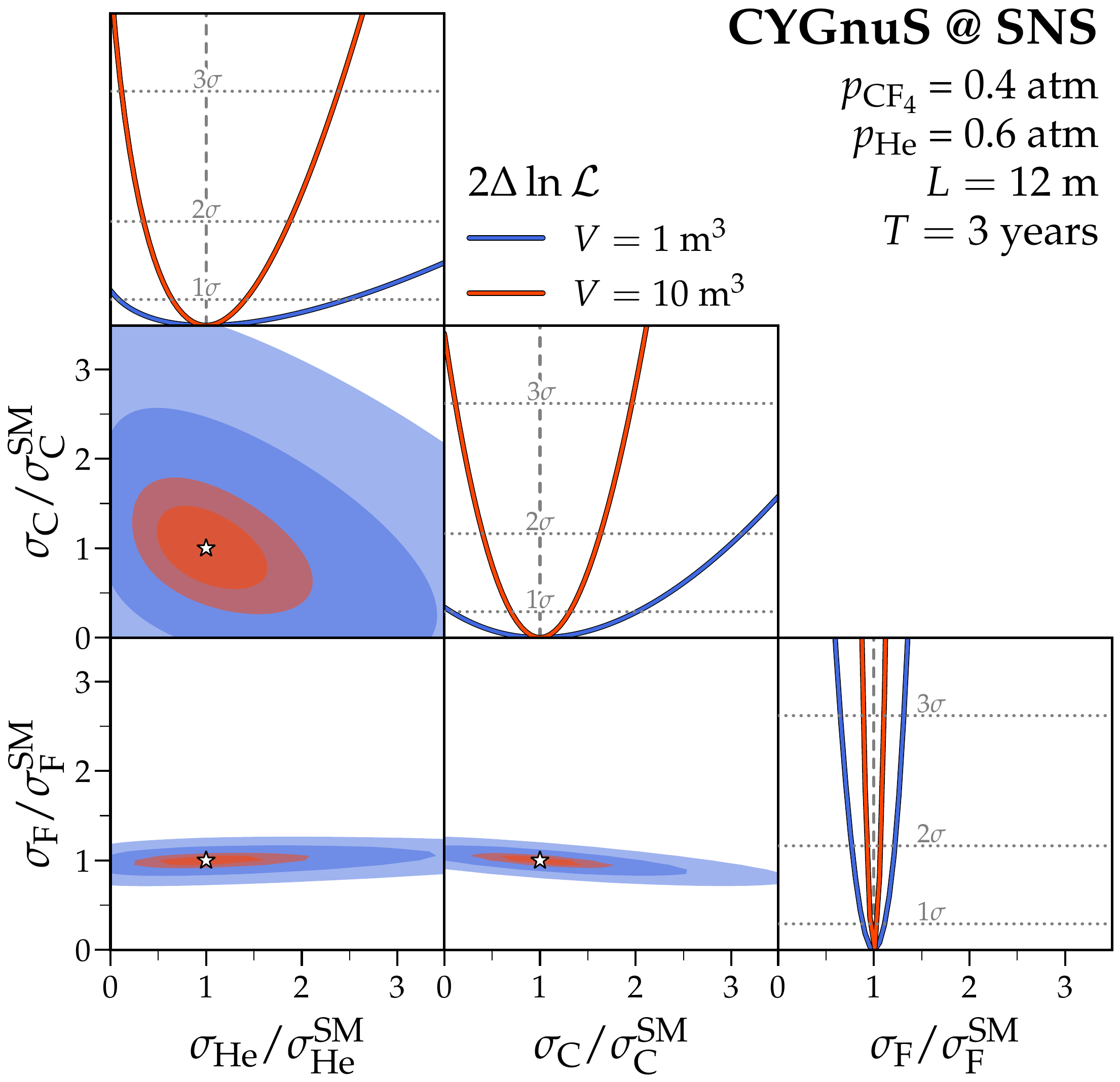}
    \caption{One and two-dimensional Asimov profile likelihood ratios for the reconstruction of the three flux-normalised \cevns cross-sections, $f_{T} = \sigma_T/\sigma_T^{\rm true}$, where $\sigma_T^{\rm true}$ is the flux-normalised cross section under the SM. Blue is used for the 1~m$^3$ volume detector and the right for 10~m$^3$. In the two-dimensional panels, the two contours are for 1 and 2$\sigma$ significance, while we use horizontal lines to label the significance values in the one-dimensional panels.}
    \label{fig:NuCrossSectionReconstruction}
\end{figure}

\begin{figure}[t]
    \centering
    \includegraphics[width=0.96\linewidth]{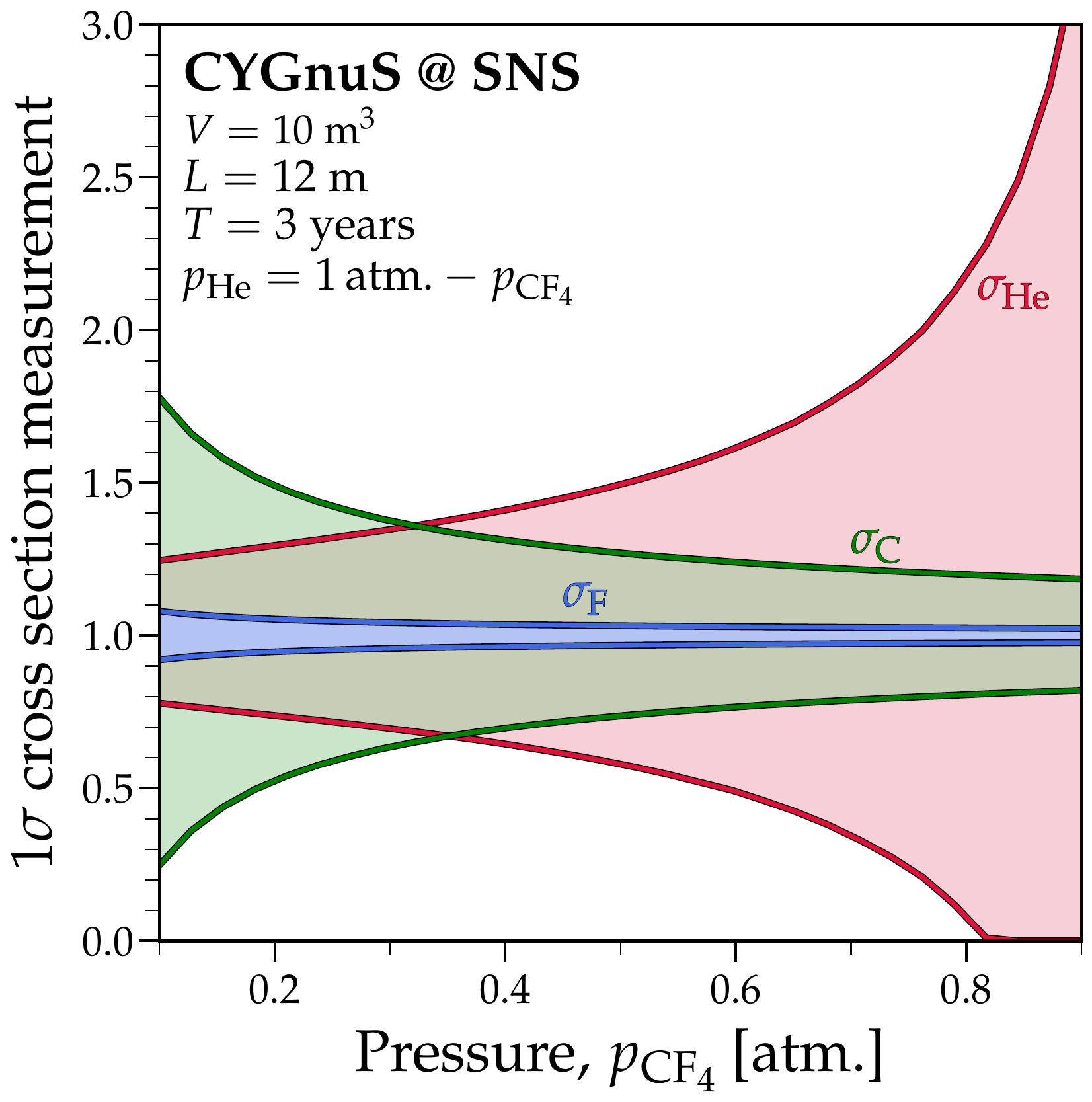}
    \caption{Median $1\sigma$ uncertainty on the reconstructed target-dependent flux-normalised cross sections as a function of the \cffour gas pressure, assuming that the helium pressure is always $p_{\rm He} = 1\,{\rm atm} - p_{\rm CF_4}$ so that the total operating pressure remains at one atmosphere. The optimal \cffour pressure for measuring all three cross sections is between 0.3 and 0.4 atm.}
    \label{fig:CrossSectionMeasurement_vs_pressure}
\end{figure}
For our first application of the statistical formalisms we have introduced, we ask the following question: assuming the Standard Model cross section, to what precision can we measure the neutrino flux? Answering this question allows us to assess the extent to which the directionality aids in both the rejection of the background and the ability to separate out the different fluxes using the available kinematic information. Recall that the SM cross section is flavour-blind, so the only way to access flavour information is through the differing energy spectra of the three fluxes.

In Fig.~\ref{fig:NeutrinoEnergyReconstruction} we show two potential flux reconstructions that we label ``model-dependent'' and ``model-independent''. In the former, we use the profile likelihood ratio test statistic and set $\boldsymbol{\theta} = \{\boldsymbol{f}_\nu,f_{\rm bg} \}$ to be our free parameters, keeping the cross-section normalisations $\boldsymbol{f}_T$ fixed at one. For the model-\textit{in}dependent approach, we use the technique described in Sec.~\ref{sec:energyreconstruction} in which we unfold the probability distributions $P(E_\nu|E_r,\cos\theta)$ for each event and then combine them to reconstruct the flux without relying on any prior knowledge of the original neutrino energy distribution. Naturally, we expect the precision of the measurement for the model-dependent case to be better, as one must supply more prior knowledge about the neutrino signal.

To construct the left-hand plot in Fig.~\ref{fig:NeutrinoEnergyReconstruction}, we compute the median 1$\sigma$ uncertainty bands around each of the best-fit fluxes (which are exactly equal to the true fluxes due to the assumption of Asimov data) for a 1~m$^3$ and 10~m$^3$ detector and assuming equal numbers of signal and background events. For the right-hand plot, however, since our model-independent approach uses unbinned rather than binned data, we cannot construct the Asimov dataset. So instead, to visualise the expected precision of the reconstruction, we show one example of the flux reconstruction using a single set of randomly generated data (points with error bars), while the blue bands contain 68 and 95\% of 10,000 simulated experiments. The binning used to construct this plot is arbitrary---we have chosen a suitable binning by eye that minimises the Poisson noise present in the reconstruction while still revealing the general expected shape of the spectrum. 

Starting first with the model-dependent case, we see that all three fluxes are inconsistent with zero for the 10~m$^3$ volume at more than 1$\sigma$, however, for the 1~m$^3$ case, only two of the fluxes could be measured, while the $\nu_e$ flux is consistent with zero. The reason for this will be made clearer by the next plot---it is due to a degeneracy between the $\nu_e$ and $\bar{\nu}_\mu$ recoil distributions, that persists even when directional information is included. 

For the model-independent reconstruction, we only show the 10~m$^3$ case because there are not enough events expected in a 1~m$^3$ experiment to provide a very meaningful constraint on the shape of the flux (i.e.~we would have to construct the flux out of very few neutrino energy bins). The expected precision of the measurement, encapsulated in the teal bands, gets noticeably poor at small neutrino energies, which we understand to be a consequence of the worsening energy and angular resolutions at low recoil energies. We also notice that non-zero values of the flux are reconstructed at energies larger than any of the true neutrino energies in the simulated data $(E_\nu \gtrsim 50~{\rm MeV})$, which again is due to the limitations in the energy reconstruction due to the energy/angular resolutions.

Next, in Fig.~\ref{fig:NuFluxReconstruction}, we show how a joint measurement of the three fluxes could be achieved using this approach. We show one and two-dimensional profile likelihood ratios for 1~m$^3$ and 10~m$^3$ experiments. In the 10~m$^3$ case, we find that a $>3\sigma$ measurement of the $\bar{\nu}_\mu$ and a $>2\sigma$ measurement of the $\nu_\mu$ flux is possible, but only $\sim 1\sigma$ significance measurement is possible for the $\nu_e$ flux. This is due to the slightly lower event rate for this latter flux and the fact that it is degenerate with the other fluxes. Similarly, the reason why the $\nu_\mu$ flux is reconstructed better than the other two is primarily thanks to its highly distinct recoil energy-angle distribution (cf.~Fig.~\ref{fig:RecoilDistributions}). Although the degeneracy between the fluxes is strong, it is already clear in the 10 m$^3$ case that a joint measurement of all three fluxes would be possible with slightly more statistics because the orange contours are already visibly beginning to close. We reiterate again that we are taking a pessimistic approach by floating the three flux normalisations (as well as the background rate) as fully free parameters in our likelihood---we do this to showcase what could be achieved with this kind of data alone. In practice, applying a $\sim$10--20\% Gaussian constraint to the total normalisation would be a very reasonable option given current knowledge of the SNS neutrino flux, and this would understandably improve all of our sensitivities beyond what we present here.

To understand the impact our particular gas mixture has on the sensitivity to the flux normalisations, we now collapse the information presented in the previous two figures down to just the median $1\sigma$ band around the model-dependent measurement of each flux, which allows us to then plot this uncertainty as a function of the CF$_4$ pressure. We show this result in Fig.~\ref{fig:FluxMeasurement_vs_pressure}. Recall that when we change $p_{\rm CF_4}$, we are adjusting the helium fraction to bring the total gas pressure up to one atmosphere (see the discussion in Sec.~\ref{sec:detector} about the advantages of using helium to retain atmospheric pressure operation). We observe that the sensitivity to all three fluxes worsens with decreasing $p_{\rm CF_4}$, which can be understood straightforwardly as a result of lower statistics, since fluorine recoils dominate the signal. That said, the scaling of the upper limit in the uncertainty bands is slightly weaker than what we would expect from statistics alone, i.e.~$1/\sqrt{p_{\rm CF_4}}$, which shows that the worsening directionality at higher gas densities is playing a role in degrading the sensitivity. The conclusion here is that a higher gas pressure of CF$_4$ is preferable in general in the context of this statistics-limited measurement, although increasing it beyond 0.4~atm does not significantly improve the precision because \cffour pressures around a full atmosphere would lead to an almost total sacrifice of all recoil directionality across most of the spectrum. The specific value of $p_{\rm CF_4} = 0.4$~atm that we have adopted for our fiducial gas mixture will be further motivated in the next section, where another trade-off presents itself.

\subsection{CEvNS cross-section}\label{sec:measure_crosssection}
We now perform a variant of the analysis presented in the previous section, but instead of fixing the SM cross-section and reconstructing the flavour-dependent fluxes, we instead float the flux-averaged values of the target-dependent cross sections to determine how precisely the \cevns process itself can be measured, and its $N^2$-dependence confirmed.

In Fig.~\ref{fig:NuCrossSectionReconstruction}, we show a comparable plot to Fig.~\ref{fig:NuFluxReconstruction} but now for the three flux-averaged cross-sections: $(\sigma_{\rm He},\sigma_{\rm C},\sigma_{\rm F})$. As before, we fix the pressure ratio to 60:40 \hecf and show two results for a 1 and 10~m$^3$ experiment running for three years and assuming equal numbers of signal and background events.

We emphasise that we are not assuming any individual recoil species identification based on the tracks themselves; the discrimination between the three target nuclei here is made possible purely due to the kinematics, i.e.~the differing recoil energy and angle distributions shown in Fig.~\ref{fig:RecoilDistributions}. We note in passing that it may be possible to identify the recoiling nucleus species at the level of individual events from measurable quantities such as the track length and the ionisation profile along the track ($\mathrm{d}E/\mathrm{d}x$). Including this information would further improve our flux and cross-section measurements, but modelling it accurately would require much more detailed gas simulations.

As we would naively expect, only the $^{19}$F cross section is measured at high significance in a 1~m$^3$ experiment because the event rates due to the other two nuclei are low. The value of $\sigma_{\rm F}$ can be measured at $>4\sigma$ in both 1 m$^3$ and 10 m$^3$. The other two cross sections are also just detectable at $3\sigma$ significance when we go to a 10 m$^3$ experiment. We emphasise that this is not achievable without directional information due to the strong degeneracy between the three cross sections driven by the similarities in their $\textrm{d}R/\textrm{d}E_r$ profiles. This degeneracy can be almost entirely relaxed when we add the $\cos\theta$ information, as can be understood by comparing their recoil distributions in Fig.~\ref{fig:RecoilDistributions} 

In a similar way to the neutrino fluxes in Fig.~\ref{fig:FluxMeasurement_vs_pressure}, we can also collapse this information down to plot the median $1\sigma$ measurement band on each cross section as a function of the \cffour pressure. We show this result in Fig.~\ref{fig:CrossSectionMeasurement_vs_pressure}. As before, the sensitivity is still driven by a mixture of statistics (which improve towards \textit{higher} $p_{\rm CF_4}$) and good directionality (which improves towards \textit{lower} $p_{\rm CF_4}$). Here, however, the trade-off has an additional factor because the measurement of the helium cross section is of course impossible if $p_{\rm He} \to 0$. So a balanced trade-off for optimising all three measurements occurs around $p_{\rm CF_4} \approx 0.4$~atm. This result provides further motivation (beyond the discussion in previous sections) for a relatively low-pressure of \cffour gas despite the sacrifice in overall event statistics that comes with using a low-density target medium.

\begin{figure}[t]
    \centering
    \includegraphics[width=0.99\linewidth]{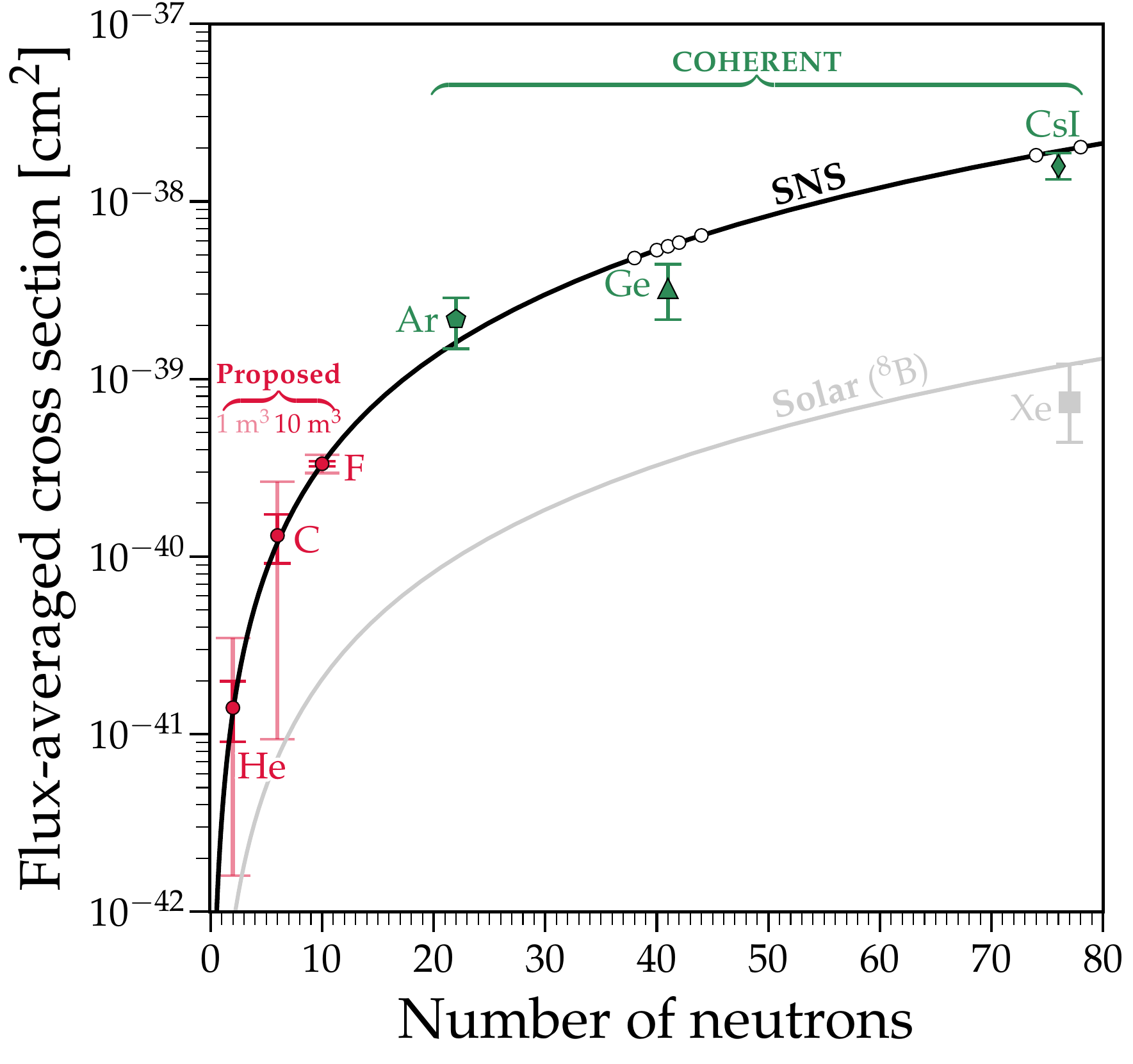}
    \caption{Projected median $1\sigma$ measurements of the total cross section for each target nucleus assuming our fiducial 60:40 \hecf gas mixture, three years of operation and equal numbers of signal and background events. The dark-red errorbars are for a 10 m$^3$ experiment and the light-red errorbars for 1 m$^3$. We point out that the median He cross-section measurement is consistent with zero at 1$\sigma$ in the 1 m$^3$ case due to the low expected event numbers. We overlay existing measurements from COHERENT at SNS~\cite{COHERENT:2022nrm,COHERENT:2020iec, COHERENT:2020ybo,COHERENT:2024axu} and solar $^8$B measurements by LZ~\cite{Akerib:2025xla} (see also XENONnT~\cite{XENON:2024ijk} and PandaX~\cite{PandaX:2024muv}).}
    \label{fig:TotalCrossSection}
\end{figure}

To finish this discussion, we summarise our three proposed cross-section measurements in Fig.~\ref{fig:TotalCrossSection}. This plot is inspired by similar ones shown by COHERENT in e.g.~Refs.~\cite{Adhikari:2026qrv}, although we caution that information about the uncertain nuclear form factor and the differing energy thresholds of different experiments is implicitly folded in---the plot should be taken as primarily illustrative, and a consistency check on the expected $\sim N^2$ scaling. We show the three measurements for both 1 and 10~m$^3$ scale experiments (light and dark red respectively), comparing them with the existing measurements from COHERENT at SNS; and liquid-xenon detectors measuring $^8$B solar neutrinos. The dependence of the flux-averaged cross section on the neutrino's energy spectrum is the reason why the solar $^8$B measurements lie along a different line to the measurements taking place at the SNS. This plot highlights the complementary nature of our proposal in the context of the global \cevns program.

\subsection{The Migdal effect}\label{sec:Migdal}

\begin{figure*}
    \centering
    \includegraphics[height=0.47\linewidth]{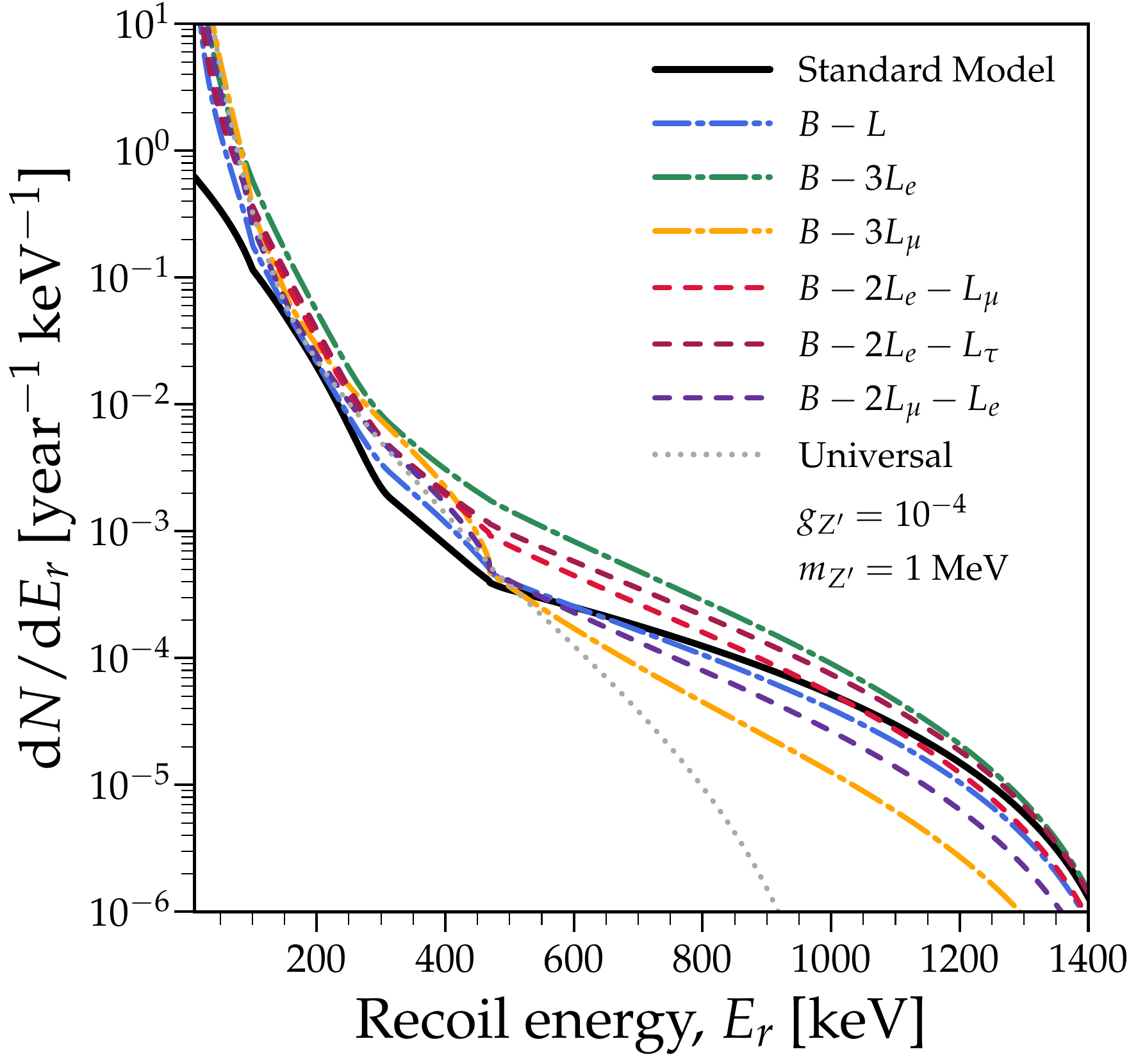}
    \includegraphics[height=0.463\linewidth]{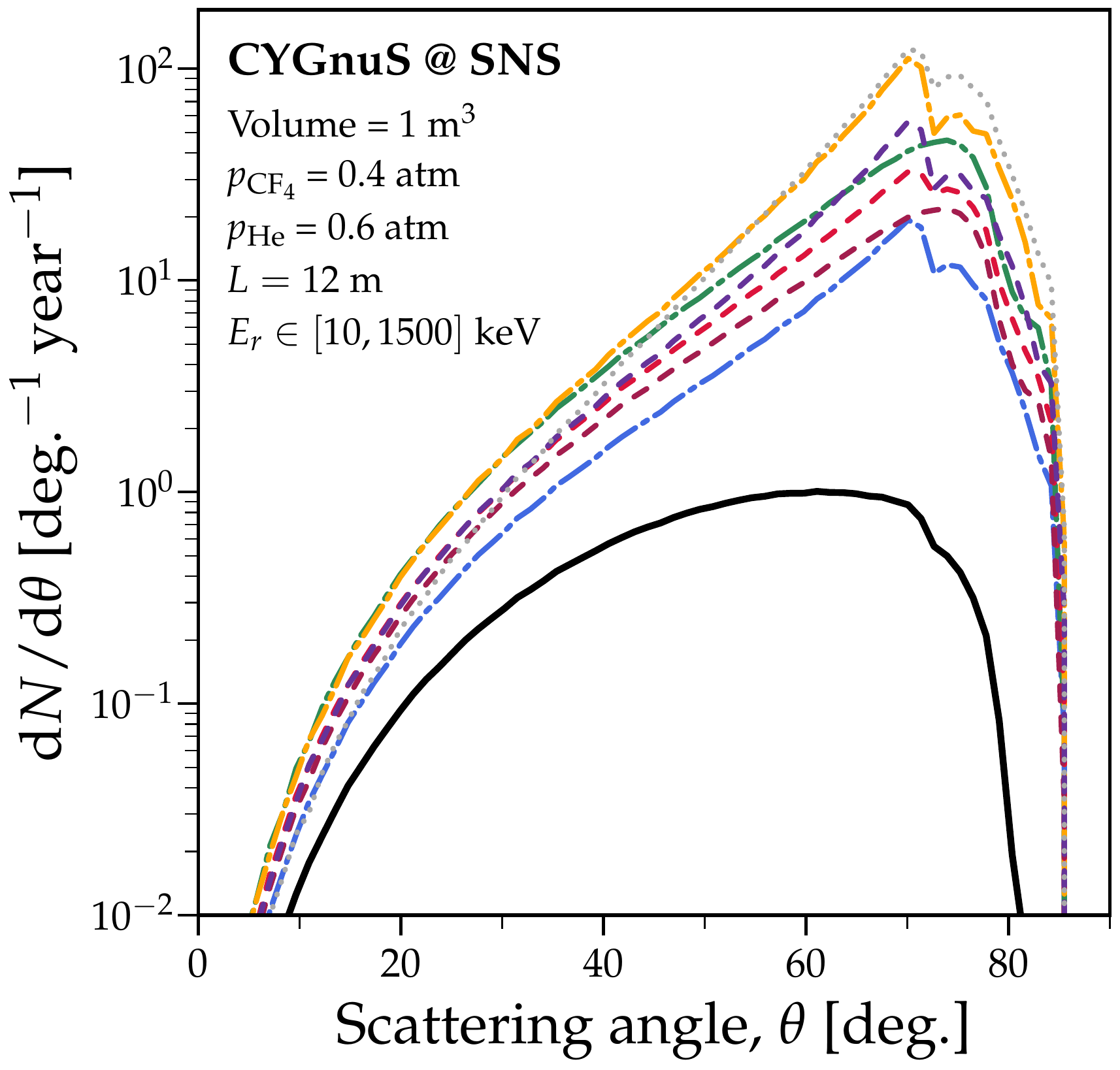}
    \caption{{\bf Left:} Total event rates summed over all three neutrino fluxes and the three target nuclei in our fiducial gas mixture (60:40 \hecf), for various low-mass vector mediator models. We show all anomaly-free models listed in Table~\ref{tab:U(1) charges} as well as the case of a universally-coupled $Z^\prime$. The SM case is shown by a solid black line for comparison. The mediator mass and coupling are fixed to $m_{Z^\prime} = 1$~MeV and $g_{Z^\prime} = 10^{-4}$ in all cases so that the effects of the additional interactions on the rate are visible. {\bf Right}: Angular distributions for the same set of models, obtained by integrating the event rate over the energy window $E_r \in [10,1500]\,{\rm keV}$.}
    \label{fig:VectorMediator_EventRates}
\end{figure*}
The final SM physics case that could be of interest to this type of experiment is a measurement of the neutrino-induced Migdal effect. This effect occurs when the sudden displacement of a nucleus leads to the emission of one or more atomic electrons~\cite{Migdal:1939, Feinberg:1941}. The Migdal effect has picked up interest recently because it provides another channel for recoil-based searches for dark matter to observe signal events, because the emitted electron could be seen even when the nuclear recoil energy falls below threshold, thereby potentially extending the sensitivity of experiments to the sub-GeV mass range~\cite{Ibe:2017yqa, Dolan:2017xbu}. The use of the Migdal effect as a signature for neutrinos has also been proposed as a novel way to probe the \cevns signal~\cite{Ibe:2017yqa, Bell:2019egg, Maity:2024hzb}.

Despite this resurgence in interest, the Migdal effect remains to be definitively calibrated in the low-energy recoil regime. A measurement of this effect in neutron-induced nuclear recoils is also being pursued by several collaborations~\cite{Xu:2023wev}, including MIGDAL~\cite{MIGDAL:2022yip, Tilly:2023fhw, MIGDAL:2024alc}, who employ similar detector technology as we are proposing here. Although a recent direct observation was reported using high-energy D-D neutrons~\cite{Yi:2026fmf}, experimental validation in the kinematic regime relevant for dark matter is lacking. Previous work studied the prospects of measuring the effect using the neutrino flux at the SNS~\cite{Bell:2021ihi}, but the setup was found to be untenable, primarily due to the difficulty of separating the signal from the background of nuclear recoils in liquid-phase detectors. However, the use of direction-sensitive gaseous detectors offers the added advantage of topological identification and reconstruction of the simultaneous nuclear and electron recoil tracks~\cite{Nakamura:2020kex}.

The signal of a Migdal event would therefore be a nuclear recoil track and a low-energy electron recoil track whose heads would be connected by a common vertex. To assess the viability of seeing this effect at the SNS, we calculate the rate of Migdal events using the formalism of \cite{Ibe:2017yqa} and the more accurate ionisation probabilities from \cite{Cox:2022ekg}. We compute the number of single ionisation events, where a single electron is ejected from the recoiling atom. Assuming our fiducial gas mixture, and a nuclear recoil threshold of 10~keV$_r$, we find the total rate of Migdal events to be 6.1~events/m$^3$/year, with the vast majority ($>$87\%) occurring due to scattering on fluorine. Such events are not an \textit{additional} contribution to the \cevns rate, instead they constitute a modified \cevns signal morphology (i.e.~$\sim$6 of the $\sim$37 \cevns events per year in a m$^3$ experiment would be Migdal events). 

We note that this treatment uses the isolated atom approximation and thus neglects molecular effects, which could alter transition probabilities and induce an anisotropy~\cite{Blanco:2022pkt}. However, the nuclear energies considered here are much greater than the molecular binding energies, meaning we can approximately treat the atoms as free. A detailed treatment of the molecular effects is beyond the scope of the present work, but may provide interesting modifications to the angular distribution of Migdal events.

This event rate is potentially observable with a 10~m$^3$ detector; however, experimental identification is unfortunately complicated by the fact that most emitted electrons have very small energies ($\lesssim$~50 eV) and thus are unlikely to leave easily discernible tracks. Nevertheless, as the scattered ion is frequently left in an excited state (if the electron is ejected from an inner shell), there is a possibility of identifying events via the accompanying signal from Auger electrons and/or fluorescence photons. We defer a more detailed analysis of the complete morphology of the Migdal signal and its identification to a future study.

\section{Beyond-Standard Model measurements}\label{sec:BSM}
Having determined our expected sensitivity to measurements in the context of the SM and used them to optimise a gas mixture, we will now explore the potential to probe interactions of neutrinos beyond the SM. We will focus on two main science cases which are popular in the \cevns literature, namely testing beyond-SM mediators participating in the neutrino-nucleus interaction, as well as the existence of a light sterile neutrino.

\subsection{New mediators}\label{sec:lightmediators}
\begin{figure}
    \centering
    \includegraphics[width=0.99\linewidth]{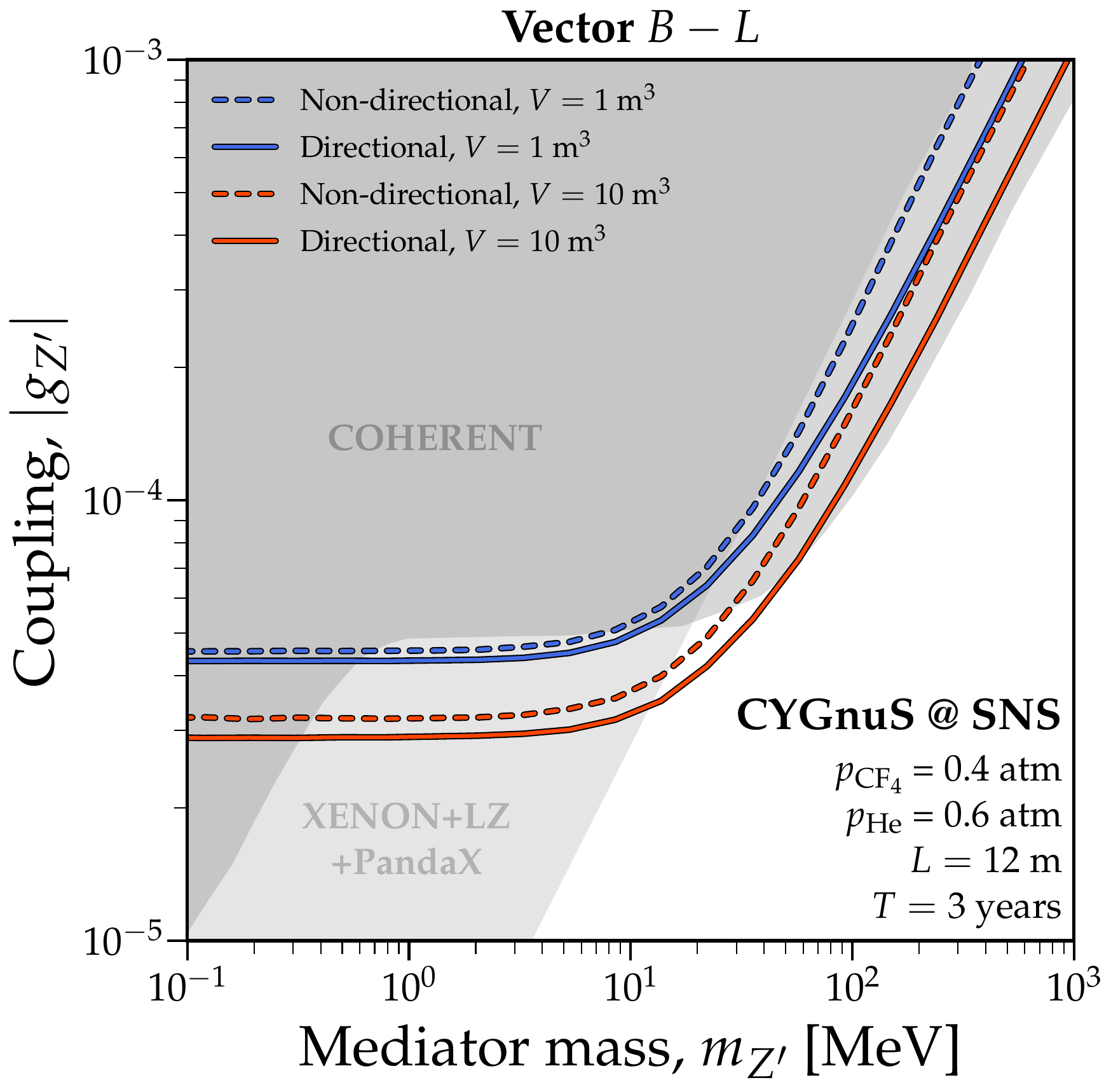}
    \caption{Projected median 95\% CL upper limits on a vector $B-L$ model for 1 m$^3$ (blue) and 10 m$^3$ (orange) experiment. For comparison, the dashed lines show the limit obtained when we repeat the same analysis while ignoring all directionality and only incorporating recoil energy information. These projections are to be compared with existing constraints estimated from COHERENT (CsI+LAr data combined) and liquid-xenon dark matter detectors (XENON, PandaX and LZ data combined) presented in Refs.~\cite{DeRomeri:2022twg, DeRomeri:2024dbv}. There are comparable constraints on this model from cosmology, stellar cooling and other neutrino oscillation/scattering measurements, see e.g.~Ref~\cite{DeRomeri:2024dbv}.}
    \label{fig:VectorMediator_constraints}
\end{figure}
\begin{figure*}
    \centering
    \includegraphics[width=0.92\linewidth]{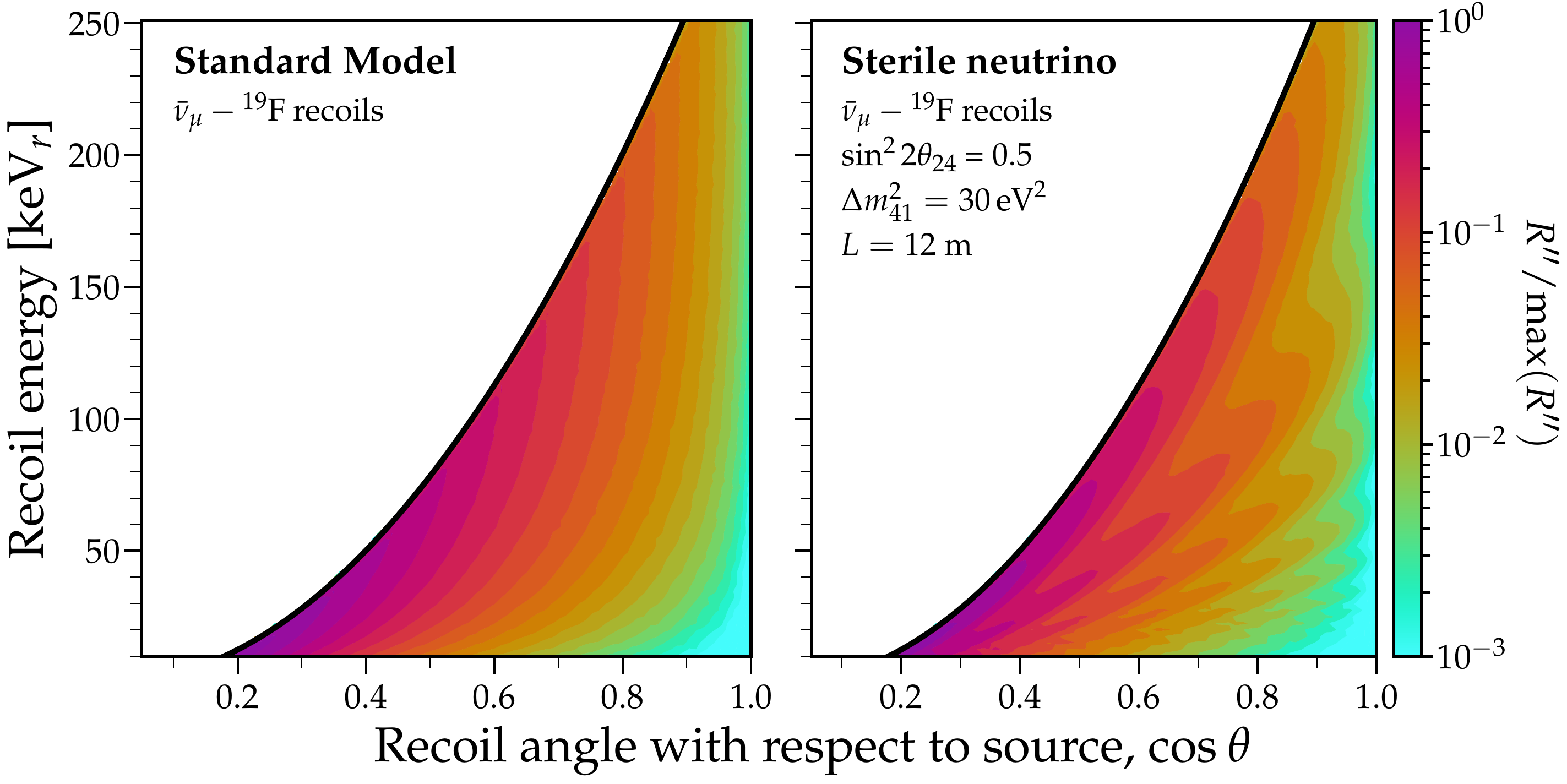}
    \caption{Normalised recoil energy and angle distributions for $\bar{\nu}_\mu$ neutrinos scattering on $^{19}$F. The left-hand plot shows the Standard Model recoil distribution for comparison, while the right-hand plot shows the case when a sterile neutrino is present, taking the excluded values $\sin^2{2\theta_{24}} = 0.5$ and $\Delta m^2_{14} = 30$~eV$^2$ solely for illustrative purposes. The black line shows the maximum recoil energy for a given scattering angle allowed by the kinematic constraint.}
    \label{fig:SterileNu_RecoilDistributions}
\end{figure*}

We consider the existence of a new lepton-flavour-conserving $U(1)$ vector gauge boson $Z'$ coupled to neutrinos and quarks. The Lagrangian describing this new interaction can be written as~\cite{AristizabalSierra:2019ykk,Abdullah:2020iiv,Cadeddu:2020nbr,DeRomeri:2024dbv}:
\begin{equation}
    \mathcal{L} \subset g_{Z'} \left( \sum_\alpha Q_{Z'}^{\nu_\alpha} \bar{\nu}_{\alpha L} \gamma^\mu \nu_{\alpha L} + \sum_f Q_{Z'}^f\bar{f} \gamma^\mu f \right) Z'_\mu \, ,
    \label{eq: Vector NSI Lagrangian}
\end{equation}
where $Q_{Z'}^{\nu_\alpha}$ and $Q_{Z'}^f$ denote the individual model-dependent vector charges of the neutrino flavour state $\alpha$ and fermion $f$; and $g_{Z'}$ is a dimensionless coupling constant. To ensure the theory remains anomaly-free when the $U(1)$ symmetry is gauged, only certain combinations of charges are permitted. Some example models discussed in the literature are listed in Table~\ref{tab:U(1) charges} along with their corresponding quark and lepton charges.
\begin{table}[t]
\centering
\begin{tabular}{lcccc}
\hline
\hline
\textbf{Model} & $Q^{e/\nu_e}_{Z'}$ & $Q^{\mu/\nu_\mu}_{Z'}$ & $Q^{\tau/\nu_\tau}_{Z'}$ & $Q^{u/d}_{Z'}$ \\
\hline
$B - L$ & $-1$ & $-1$ & $-1$ & $1/3$ \\
$B - 3L_e$ & $-3$ & $0$ & $0$ & $1/3$ \\
$B - 3L_\mu$ & $0$ & $-3$ & $0$ & $1/3$ \\
$B - 2L_e -L_\mu$ & $-2$ & $-1$ & $0$ & $1/3$      \\
$B - 2L_e - L_\tau$ & $-2$ & $0$ & $-1$ & $1/3$    \\
$B - 2L_\mu - L_e$ & $-1$ & $-2$ & $0$ & $1/3$     \\
Universal & $1$ & $1$ & $1$ & $1$
\\
\hline
\hline
\end{tabular}
\caption{Vector charges for quarks and leptons within several anomaly-free models where various combinations of baryon and lepton-number symmetries are gauged. We also list a universally-coupled model for comparison.}
\label{tab:U(1) charges}
\end{table}
Assuming for concreteness that the vector coupling is the same for both $u$ and $d$ quarks, the nucleus' vector charge, $g_V$, in the \cevns cross-section [Eq.~\eqref{eq: SM CEVNS CS}] is modified in the following way,
\begin{equation}
    g_V \to g_V + \frac{3 g_{Z'}^2 Q_{Z'}^f Q_{Z'}^{\nu_\alpha} (Z+N)}{\sqrt{2} G_F (2m_NE_r + m_{Z'}^2)} \, ,
    \label{Vector NSI CS}
\end{equation}
where $m_{Z'}$ is the mass of the $Z^\prime$. Because this additional term can take a positive or a negative sign, there is the possibility for the new interaction to lead to both constructive and destructive interference with the SM contribution at the level of the event rate.

An alternative model choice could be to consider a scalar mediator $\phi$, with the Lagrangian~\cite{Farzan:2018gtr, AristizabalSierra:2019ykk, Abdullah:2020iiv, DeRomeri:2024iaw},
\begin{equation}
    \mathcal{L}\subset \left( g_\nu \bar{\nu}\nu + \sum_{q = u,d} g_q \bar{q}q  \right) \phi \, .
    \label{eq: Scalar NSI Lagrangian}
\end{equation}
where $g_\nu$ is the coupling of the scalar to neutrinos, and $g_q$ is the scalar coupling to quarks. The cross-section in this case is modified in the following way,
\begin{equation}
   \frac{\drm \sigma}{\drm E_r} \to \frac{\drm \sigma}{\drm E_r} + \frac{C_\phi^4 }{2\pi(2m_NE_r + m_\phi^2)^2} \left( \frac{m_N^2 E_r}{2 E_\nu^2} \right) \, , 
    \label{eq: Scalar NSI CS}
\end{equation}
where $m_\phi$ is the scalar's mass. Considering the case of universal coupling where the mediators couple to neutrinos and quarks with the same strength, we can write the resulting cross section in terms of one coupling parameter $g_\phi^2 = g_\nu g_q$. Then, $C_\phi$ in Eq.\eqref{eq: Scalar NSI CS} is given by, 
\begin{equation}
    C_\phi^2 = g_\phi^2 \left( Z \sum_{q = u,d} \frac{m_p}{m_q} f_q^p + N \sum_{q = u,d} \frac{m_n}{m_q} f_q^n \right) \, ,
    \label{eq: Scalar Quark Coupling}
\end{equation}
where $m_p$ and $m_n$ are the masses of the proton and neutron; $m_q$ are the quark masses; $f_q^p$ and $f_q^n$ are the scalar form factors in protons and neutrons. Unlike the vector case, there is no interference in the cross section for this model, so the new interaction simply enhances the rate at low energies. This makes the scalar mediator somewhat less interesting for the purposes of this study, given that we merely wish to highlight non-standard neutrino interactions as a potential physics case, as opposed to exhaustively forecasting sensitivity to every iteration of this type of model. For this reason, we will show results only for the vector case and leave a full exploration of this detector's ability to test non-standard neutrino interactions for a future study, once more experimental details have been decided.

We show the recoil energy and recoil angle distributions for a set of light vector mediator models in Fig.~\ref{fig:VectorMediator_EventRates}, assuming a benchmark mediator mass of $m_{Z^\prime} = 1$~MeV. These models modify the recoil spectrum and generally lead to an enhancement in the rate towards low energies. There is also a corresponding shift in the angular spectrum towards larger scattering angles because of this same enhancement at low $E_r$. Qualitatively, the effect of adding these new interactions is similar across all models, so we will show only the sensitivity to the $B-L$ model, as this is the example that demonstrates the best complementarity with COHERENT. We have checked that the forecasted sensitivities to all models are comparable.

Assuming that the measured rate is consistent with the Standard Model, we can deploy the same likelihood formalism to project 95\% CL upper limits on the value of $g_{Z^\prime}$ given a fixed value of $m_{Z^\prime}$. Following the convention for these kinds of constraints, we then project the mass-dependent limit by repeating the test over a range of fixed $m_{Z^\prime}$ values. The one-sided 95\% CL \textit{exclusion} limit on $|g_{Z^\prime}|$ is then drawn when $2\Delta \ln \mathcal{L} = -2.71$. Fig.~\ref{fig:VectorMediator_constraints} shows the projected limit for 1 and 10~m$^3$ volumes, assuming three years of operation and equal numbers of background and signal events. 

To illustrate the fact that the directionality provides additional statistical power here (in terms of both background rejection and in discriminating the SM and BSM recoil distributions), we repeat the analysis while ignoring the $|\cos{\theta}|$ observable---i.e. a ``non-directional'' experiment that otherwise sees an identical event rate.\footnote{In practice, this simply amounts to summing the rate over the $\cos\theta$ bins before calculating the likelihood}. We forecast a sensitivity that is competitive with the existing COHERENT and liquid-xenon-based experimental bounds. The use of light target nuclei here gives these constraints a slightly different shape COHERENT's bound, noticeable most in the small window around $m_{Z^\prime} \sim 20 $~MeV, where the projected limits extend into a region not excluded by that experiment.

\begin{figure}
        \centering
        \includegraphics[width=0.95\linewidth]{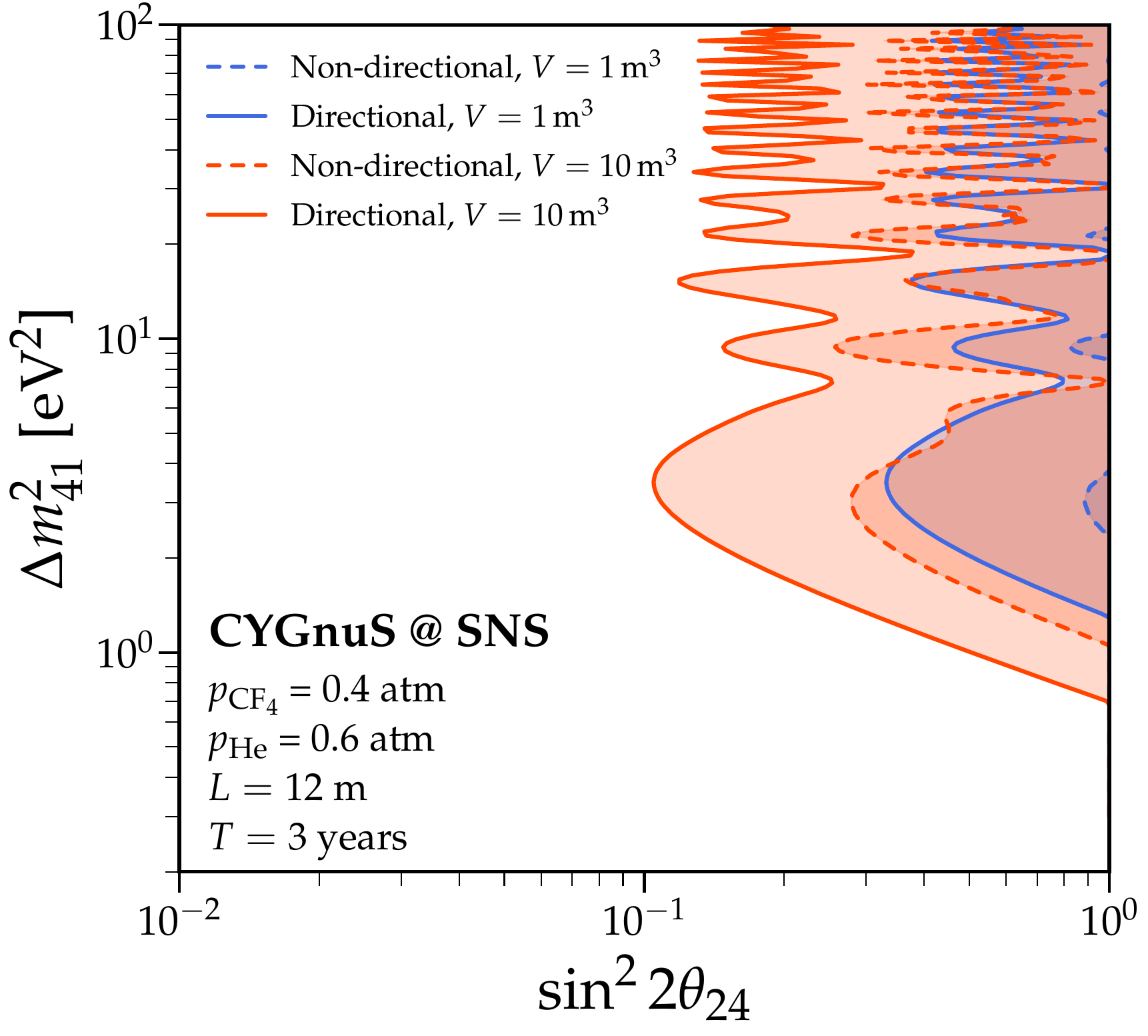}
        \caption{Projected median 90\% CL exclusion limits for a 1 and 10 m$^3$ experiment, on the sterile neutrino mass difference and mixing angles, showing just the $\theta_{24}$ case for illustrative purposes. As in previous figures, the blue lines are for a 1~m$^3$ experiment, while the orange lines are for a 10~m$^3$ experiment. We also highlight the role played by the directional information by repeating the analysis while ignoring the $\cos\theta$ observable and only using $E_r$--these results are shown by dashed lines.}
        \label{fig:SterileNu_projection}
\end{figure}
\subsection{Sterile neutrinos}\label{sec:sterileneutrinos}
We finally come to the case of extending the SM to include a sterile neutrino with $\gtrsim$eV-scale mass. The existence of a new non-interacting neutrino flavour state is expected to affect the spectrum of the three interacting flavours of neutrinos arriving at the detector, which would then be imprinted on the recoil distribution.

In the short-baseline limit, the transition to different flavours is suppressed if $\Delta m^2 L / 4 E_\nu \ll 1$. As a result, in the standard three-flavour framework, where $\Delta m^2\lesssim10^{-3}$~eV$^2$, no flavour oscillations are observable on scales below $\sim100$ m for neutrinos with $E_\nu \gtrsim 1$ MeV. However, the possible existence of a light sterile neutrino modifies the $3\times3$ leptonic mixing matrix (U) into a $4\times4$ matrix, and adds a new mass-squared difference, $\Delta m_{41}^2$. For large values of $\Delta m^2_{41}\gtrsim 1$~eV$^2$, oscillations can occur on short, 1--10~m scales. For this scenario, the transition probabilities are given by;
\begin{align*}
    \text{P}_{es}(E_\nu) &= 4 |\text{U}_{e4}|^2 |\text{U}_{s4}|^2 \sin ^2 \left(\frac{\Delta m^2_{41} L}{4 E_\nu}\right) \\
    &= \sin^2 2\theta_{14} \cos^2 \theta_{24} \cos^2 \theta_{34} \sin ^2 \left(\frac{\Delta m^2_{41} L}{4 E_\nu}\right)\, , \\
    \text{P}_{\mu s}(E_\nu) &= 4 |\text{U}_{\mu4}|^2 |\text{U}_{s4}|^2 \sin ^2 \left(\frac{\Delta m^2_{41} L}{4 E_\nu}\right) \\
    &= \cos^4 \theta_{14} \sin^2 2\theta_{24} \cos^2 \theta_{34} \sin ^2 \left(\frac{\Delta m^2_{41} L}{4 E_\nu}\right).
\end{align*}
To incorporate the sterile neutrino into our rate calculation, we simply multiply each neutrino flux by its corresponding survival probability, i.e.~$\left(1-\text{P}_{es}\right)$ for the electron neutrino flux and $\left(1-\text{P}_{\mu s}\right)$ for the muon and anti-muon neutrino fluxes.

While this scenario is constrained by a variety of oscillation experiments~\cite{Dentler:2018sju,MicroBooNE:2025nll,Lister:2026jab}, \cevns experiments provide a unique and complementary probe~\cite{Blanco:2019vyp,Bisset:2023oxt,Chattaraj:2026htz}. Unfortunately, since \cevns experiments currently only measure nuclear recoil energies, they are only indirectly sensitive to the neutrino-energy-dependent survival probabilities. Because each neutrino energy generates a broad distribution of recoil energies, the effect of a sterile neutrino on $\textrm{d}R/\textrm{d}E_r$ is almost indistinguishable from a small reduction in the overall rate. In contrast, the inclusion of directional information on each recoil gives us a way to access the neutrino energy spectrum, and thus access to the energy-dependent survival probabilities, $P_{es,\mu s}(E_\nu)$, despite there not being any flavour information present at the level of individual recoils. This effect can be seen in the right-hand panel of Fig.~\ref{fig:SterileNu_RecoilDistributions}, where the sterile oscillations show up in the iso-rate contours of the recoil energy versus recoil angle distribution. If we integrated these distributions over either $\cos\theta$ or $E_r$ the effect would largely disappear, emphasising the importance of measuring both quantities.

To illustrate the possible sensitivity to this kind of model, we project median 90\% CL exclusion limits on $\sin^2(2\theta_{24})$ as a function of $\Delta m^2_{41}$ for 1 and 10~m$^3$ volume TPCs operating for three years with equal numbers of signal and background events. The case for $\sin^2(2\theta_{14})$ is qualitatively similar to what we present here, but the projections are significantly weaker because of the lower event rate due to electron-neutrinos. To cast these limits, we have adopted the same likelihood-ratio test statistic as we used to project exclusion limits for $g_{Z^\prime}$ as a function of $m_{Z^\prime}$ in the previous section. Since these are exclusion limits, we are assuming the Asimov data is described by the SM and fix $2\Delta \ln \mathcal{L} = -1.64$ for a one-sided 90\%~CL upper limit. The result of this exercise is shown in Fig.~\ref{fig:SterileNu_projection}. Although the projections are relatively modest compared to existing bounds, this result is the most striking demonstration of the power of directionality (comparing the solid and dashed lines of the same colour in Fig.~\ref{fig:SterileNu_projection}). Because of the reasons explained above, we find that the inclusion of directional information would improve the limit by up to a factor of $\sim4$ when compared to a recoil-energy-only \cevns experiment, which would struggle to differentiate the presence of a sterile neutrino from a shift in the overall event rate. 



\section{Conclusions}\label{sec:conc}
We have proposed a new class of detector that could be installed at the Spallation Neutron Source, which has the promise to complement the ongoing program of \cevns measurements being made by the COHERENT collaboration. Our study aligns with previous studies of the physics case of MPGD-based low-energy recoil detectors for neutrino physics in Refs.~\cite{Abdullah:2020iiv,Lisotti:2024fco,Shekar:2025xhx}---namely that the ability to simultaneously reconstruct recoil track directions as well as recoil energies enables novel approaches to various physics measurements thanks to the added signal information and background rejection that comes via directionality. We have illustrated this in the context of the purely Standard Model measurements (Sec.~\ref{sec:SM}) and have explored several options for searching for new physics beyond the Standard Model (Sec.~\ref{sec:BSM}).

We have forecasted the sensitivity for two sizes of experiment: a 1~m$^3$ path-finding experiment that is large enough to make an initial measurement, and a full-scale 10~m$^3$ volume TPC, which would have high-enough statistics for physics. We believe both experiments are feasible at the SNS site and could be positioned at a relatively nearby distance-to-source of $L = 12$~m while accommodating all necessary shielding. Our main results are shown in Figs.~\ref{fig:NuFluxReconstruction} and \ref{fig:NuCrossSectionReconstruction} for measurements within the SM, namely the measurement of the flavour-dependent neutrino fluxes, and the flux-averaged cross sections for different target nuclei. As well as providing directionality, our proposal is novel with respect to existing experiments in that it would incorporate the lightest target nuclei used to date for measuring \cevns. 

Our key result for testing beyond-SM interactions is shown in Fig.~\ref{fig:VectorMediator_constraints}, and for testing the presence of a $\mathcal{O}(10\,{\rm eV})$-mass sterile neutrino in Fig.~\ref{fig:SterileNu_projection}. We obtain a competitive level of sensitivity despite the sacrifice in the total event rate due to the gas-based target medium and the use of light nuclei. This again is thanks to the fact that a direction-sensitive experiment has superior signal characterisation and background rejection capabilities, and comes with an intrinsic ability to access the original neutrino energy at the event level, which is impossible in all current \cevns experiments. This is made possible here because the independent measurements of the neutrino-induced recoil energy and scattering angle provide complete kinematic information, enabling the reconstruction of individual neutrino energies. If background conditions prove favourable in a real experiment, it may also be possible to empirically measure the neutrino flux event-by-event, as illustrated in Fig.~\ref{fig:NeutrinoSpectrumReconstruction}. A full assessment of the background conditions and shielding requirements for a gas-based TPC situated around 12 metres from the neutrino source will be conducted in the near future.

We have estimated the energy and angular resolutions of our envisioned detector using limited gas-simulation results. Using these results, we have found that a 60:40 ratio mixture of helium and CF$_4$ is a balanced option where we neither give up on event statistics nor directional sensitivity, which scale in opposite directions with increased gas density. However, our results remain to be validated by a full engineering study and Monte Carlo simulations of nuclear recoils in the gas mixture we propose here. Nonetheless, our results suggest a promising future for this technology at a neutrino source.\\

\section*{Acknowledgments}
CAJO and CL are supported by the Australian Research Council under the grant numbers DE220100225 and CE200100008. JLN is supported by the Australian Research Council under the grant CE200100008. LES, NM, and ACS are supported by the DOE Grant No. DE-SC0010813. ML and SEV are supported by the U.S. Department of Energy (DOE) via Award Number DE-SC0010504.

\maketitle
\flushbottom

\bibliographystyle{bibi}
\bibliography{biblio}

\end{document}